\documentclass[twocolumn,%otv not in prb secnumarabic,
amsmath,tightenlines,showpacs,superscriptaddress,aps,prb]{revtex4-1}

\usepackage{times}
\usepackage{color}
\usepackage{graphicx}% Include figure files
\usepackage{dcolumn}% Align table columns on decimal point
\usepackage{amssymb}
\usepackage{amsmath}
\usepackage{amsfonts}
\usepackage{docs}%
\usepackage{bm}%
\usepackage{sidecap}
\usepackage[colorlinks=true,linkcolor=blue,pagecolor=blue,filecolor=blue,menucolor=blue,urlcolor=blue,citecolor=blue,anchorcolor=blue]{hyperref}%
\newcommand{\etal}{\emph{et al.}}
\newcommand{\sns}{$SNS $ }
\newcommand{\sfs}{$SFS $ }

\newcommand{\sff}{$SFF $ }

\newcommand{\fs}{$S$$F $ }
\newcommand{\f}{$F $ }

\newcommand{\he}{h/\varepsilon_F}

\begin{document}

\title{ Zero Energy Peak and Triplet Correlations in  Nanoscale \textit{\textbf{SFF}} Spin-Valves} %khr

\author{Mohammad Alidoust }
\email{phymalidoust@gmail.com}
\affiliation{Department of Physics,
University of Basel, Klingelbergstrasse 82, CH-4056 Basel, Switzerland}
\author{Klaus Halterman}
\email{klaus.halterman@navy.mil} \affiliation{Michelson
Lab, Physics Division, Naval Air Warfare Center, China Lake,
California 93555, USA}
\author{Oriol T. Valls}
\email{otvalls@umn.edu}
\altaffiliation{Also at Minnesota
Supercomputer Institute, University of Minnesota, Minneapolis,
Minnesota 55455, USA}
\affiliation{School of Physics and Astronomy,
University of Minnesota, Minneapolis, Minnesota 55455, USA}

\date{\today}

\begin{abstract} 
Using a self-consistent Bogoliubov-de Gennes approach, 
we theoretically study
the proximity-induced density of states
(DOS) in  clean \sff spin-valves with noncollinear exchange fields. 
Our results clearly demonstrate a direct correlation between 
the presence of  a zero energy
peak (ZEP) in the DOS spectrum 
and the persistence  of spin-1 triplet pair correlations.
By systematically varying the geometrical and material parameters governing
the spin-valve, we point out to experimentally optimal 
system configurations where the ZEPs are most pronounced, and 
which can be effectively probed
via  scanning tunneling microscopy. 
We %also  %khx
complement these findings in the 
ballistic regime by 
employing
the Usadel formalism in the full proximity limit to investigate 
their diffusive \sff
counterparts. We determine 
the optimal normalized ferromagnetic layer
thicknesses which result in the largest ZEPs. Our results can 
serve as
guidelines in designing samples for future experiments.
\end{abstract}


\maketitle

\section{introduction}\label{sec:introduction}
The interplay of ferromagnetism and superconductivity 
 in  hybrid 
superconductor ($S$) 
ferromagnet ($F$) 
structures ($S/F$ structures) constitutes a
controllable system
in which  to study fundamental physics, including prominently that of
competing multiple broken symmetries.\cite{buzdin_rmp,bergeret_rmp} 
%For instance,
The proximity of a conventional $s$-wave superconductor with
non-aligned ferromagnetic layers, or 
a textured ferromagnet, induces both spin-singlet and odd
frequency\cite{berezi} (or equivalently odd-time\cite{halter_trip}) 
spin-triplet
correlations with $0$ and $\pm 1$ spin projections along a spin quantization
axis. %near the \fs interface, 
These triplet pairs
stem from broken  time
reversal and translations\cite{buzdin_rmp,bergeret_rmp} symmetries.
This kind of  spin-triplet pairings
originally suggested as a possible
pairing mechanism in $^3$He,~\cite{berezi} 
% and this type of pairing 
has reportedly been observed in  intermetallic compounds
such as ${\rm Sr}_{2}{\rm RuO}_4$.~\cite{Maeno,mineev} 
\fs heterostructures 
are particularly simple and  feasible experimental  systems  which
allow for direct studies
of the intrinsic
behavior of differing superconducting pairings. 
Unlike  the opposite-spin correlations, spin-1 pairing correlations
are rather insensitive to the pair-breaking effects of 
ferromagnetic exchange splitting, and hence to
the thickness of the magnetic
layers, temperature, and  magnetic scattering impurities.
The amplitudes of the opposite-spin correlations
pervading the adjacent ferromagnet,  %khx
undergo damped oscillations as a function
of position which reveals itself
in $0$-$\pi$ transitions of the 
supercurrent.~\cite{trnsp_SF2_ex,trnsp_SF1_ex,buzdin_rmp,halter_dos,brgrt_dos} 
Since about a  decade ago, several proposals have been 
put forth to achieve attainable 
and practical platforms that isolate 
and utilize the
proximity-induced\cite{berezi,halter_trip} superconducting 
triplet correlations in  \fs hybrids.~\cite{buzdin_rmp,bergeret_rmp}

%---------------------DOS related----------------------
The signatures of the proximity-induced electronic density of 
states (DOS) in the
\f layers of these hybrid structures can reveal  the existence and 
type of superconducting correlations in the 
region.~\cite{DOS_SFS1_th,DOS_SFS2_th,fazio_dos,halter_dos,brgrt_dos}
%-------------------------------------------
% pros and cons of DOS measurements should be given here
%-------------------------------------------
One promising prospect for unambiguously
detecting  triplet correlations experimentally 
involves tunneling spectroscopy experiments
which can probe the local single particle spectra encompassing the 
proximity-induced DOS.~\cite{DOS_SF1_ex,DOS_SN1_ex,DOS_SF2_ex,DOS_SF3_ex,DOS_SN2_ex,DOS_SN3_ex,
DOS_SNS1_ex,DOS_SFS2_th,DOS_SFS1_th,DOS_SNS1_th,DOS_SNS2_th,DOS_AHL_th,
DOS_norm_SF1,DOS_norm_SF2,DOS_norm_SF3,DOS_SF4_ex,DOS_SF5_ex,DOS_SF6_ex}
Nonetheless, 
competing effects can make analysis of the results
of such a  `direct' probe of 
spin-triplet superconducting correlations problematic.
The  DOS in
\sns junctions and
 \sfs heterostructures where the magnetization pattern of the  \f layer can
be either uniform or textured (including domain wall and 
nonuniform textures, such as the spiral magnetic
structure of Holmium)
has been extensively
studied.~\cite{DOS_SFS1_th,DOS_SFS2_th,DOS_SNS1_th,DOS_SNS2_th,DOS_SNS1_ex,rob4,
rob5}
It was found that the DOS in a normal metal sandwiched between two $s$-wave
superconducting banks shows
a minigap which closes by simply tuning the superconducting 
phase differences up to the value of $\pi$.
\cite{DOS_SFS1_th,DOS_SFS2_th,DOS_SNS1_th,DOS_SNS2_th} 
In contrast, the DOS can exhibit anomalous
behavior in inhomogeneous magnetic layers.
Namely, upon modulating the superconducting phase 
difference\cite{DOS_SFS2_th,DOS_SFS1_th}  
a peak arises at zero energy,
at the center of what was a minigap. 
It was also shown that
the zero energy peak (ZEP) in the  DOS for a simple textured \sfs junction can be %khx2
maximized at a $\pi$ bias.\cite{DOS_SFS2_th,DOS_SFS1_th}
The minigap-to-peak behavior of the 
DOS at zero energy is an important signature of the  emergence of 
triplet correlations.~\cite{DOS_SFS2_th,DOS_SFS1_th}
Recently it was theoretically proposed that the minigap-to-peak 
phenomenon be leveraged 
for functionality in
device platforms such as SQUIDs, to enhance their performance and
as  ultrasensitive switching devices, including a
{\it singlet-triplet} superconducting quantum magnetometer.~\cite{DOS_AHL_th}

\begin{figure}[b!]
\includegraphics[width=6.50cm]{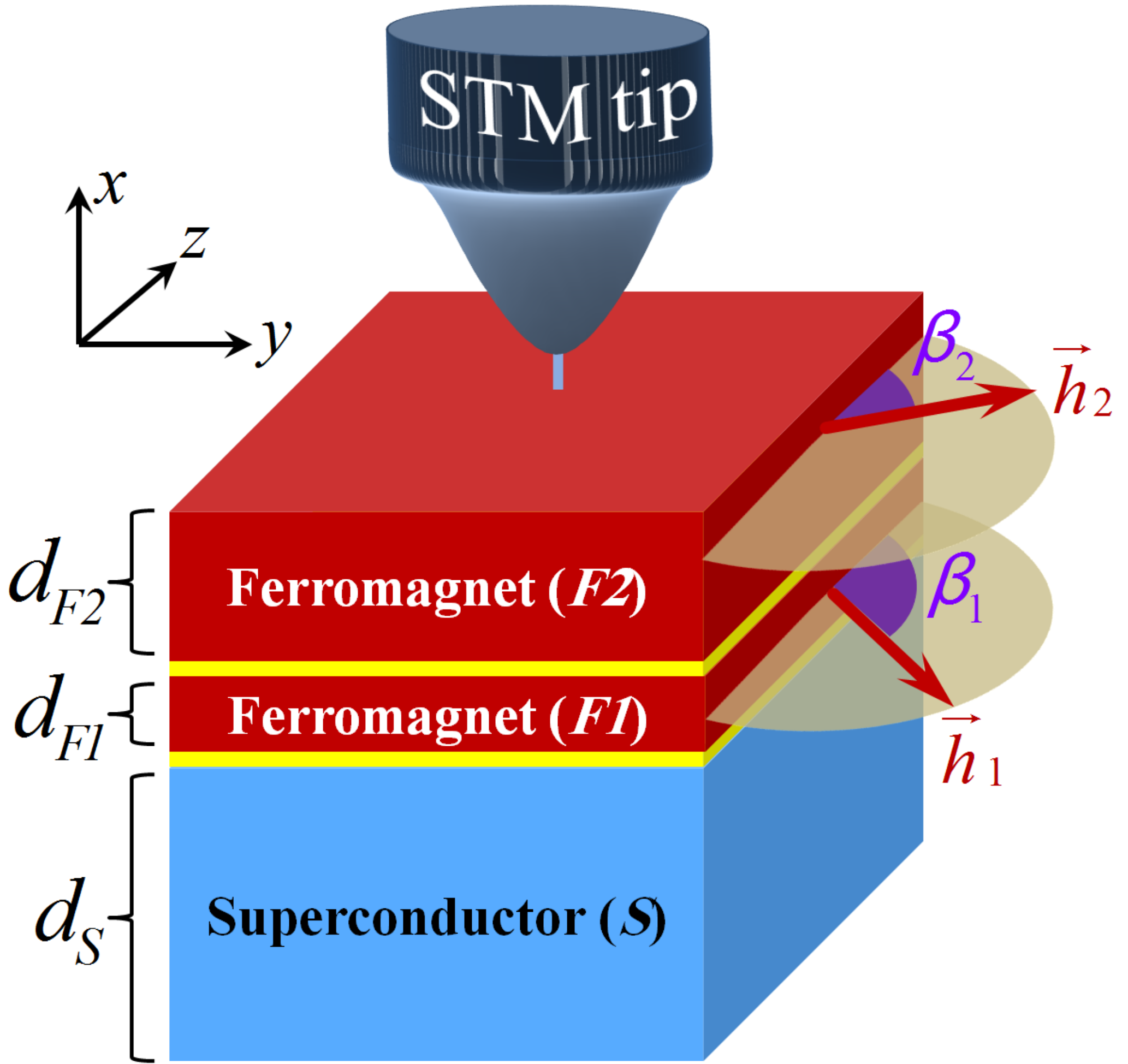}
\caption{\label{model}(Color online) 
Schematic of the \sff spin-valve structure.
The ferromagnetic layers have 
uniform exchange fields located in the $yz$ plane. %khx
The exchange field of each layer is 
defined by
$\vec{h}_{1,2}=h_0(0,\sin\beta_{1,2},\cos\beta_{1,2})$ in which $\beta_{1,2}$ are the angle of the
exchange fields with respect to the $z$ direction. The ferromagnets ($F_1$, $F_2$) and superconductor ($S$)
are stacked in the $x$ direction with thickness $d_{F1}$, $d_{F2}$, and $d_S$, respectively. The STM tip
is located at edge of the \sff spin-valve.}
\label{fig:model}
\end{figure}

An important  spectroscopic tool
for investigating  proximity effects 
on an atomic scale  with  sub-meV energy resolution
is the scanning tunneling microscope (STM).
As shown in Fig.~\ref{fig:model},
an \sff %otvx2
 spin valve structure can be probed experimentally
by positioning a nonmagnetic  STM tip at the edge of the sample
to  measure the tunneling
current ($I$) and voltage ($V$) 
characteristics. 
This technique yields a
direct probe of the available electronic
states with energy $eV$ near the tip. 
Therefore,  the
differential conductance $dI(V)/dV$ over the energy
range of interest is proportional to the
local DOS. 
Numerous experiments have 
reported signatures of the energy spectra
in this manner.~\cite{DOS_norm_SF3,DOS_SN1_ex,DOS_SF1_ex,DOS_norm_SF3,DOS_norm_SF2,
DOS_norm_SF1,DOS_SN2_ex,DOS_SF3_ex,DOS_SF5_ex,DOS_SN3_ex}
When ferromagnetic elements are present, 
the superconducting proximity-induced
DOS reveals a number of 
peculiarities due to the additional spin degree
of freedom that arises from the magnetic layers. 
However,
the experimental signatures of the
odd-frequency spin-triplet correlations 
can be  washed out
by more dominate singlet
correlations. 
When the exchange splitting $h$ of the 
magnetic layers
is  large ($\sim$$\varepsilon_F$, i.e. close to the half metallic
limit), the 
characteristic length scale %khr reword, since the f_1 component is long-range
$\xi_F$ that describes the propagation length of  %khx
opposite-spin pairs 
in the ferromagnets is extremely 
small.  
These types of proximity-induced correlations 
can thus only be experimentally observed in
weak magnetic alloys $h\ll\varepsilon_F$ (such as ${\rm Cu}_{x}{\rm Ni}_y$) or
thin $F$ layers so that $d_F$/$\xi_F$ is sufficiently  large to allow the %khx
opposite-spin superconducting correlations to
propagate in the ferromagnet without being completely suppressed.~\cite{DOS_SF1_ex,DOS_SF2_ex,DOS_SF4_ex} 
Since spin-1 triplet pairs 
are not destroyed  by the ferromagnetic exchange field
in strong magnets, 
there should exist certain 
system parameters, e.g., 
ferromagnet widths and exchange fields, 
that result in regions  whereby  
equal-spin pairs are the only pair correlations  present.
This scenario was explored in 
a $S$/Ho bilayer\cite{DOS_SF6_ex}, 
where %otvr
phase-periodic conductance oscillations were observed
in Ho wires connected to an ordinary s-wave superconductor.
This behavior was qualitatively explained
in terms of the long-range penetration of proximity-induced spin-1 triplet
pairings due to the helical structure of the magnetization.\cite{MK_jap} 
In practice however,
simpler structures involving  \fs hybrids  %khx
with uniform exchange fields %khx
are often 
preferable from both an experimental and theoretical 
perspective.~\cite{DOS_SF1_ex,DOS_SF2_ex,DOS_SF3_ex,DOS_SF4_ex,DOS_SF5_ex,DOS_SF6_ex,DOS_norm_SF3,DOS_norm_SF2,DOS_norm_SF1}
Therefore, the primary aim of this work is
the determination of experimentally optimal parameters 
for probing  odd-frequency
spin-1 triplet correlations
with DOS signatures in nanoscale \sff spin valves.

%-------------------------------------------
% Our work
%-------------------------------------------
%otvr paragraph changed 
%otvx2 edited
Nearly all of the past theoretical
works on \sff structures have considered 
the diffusive case,\cite{beasley,fominov,karmin} where impurities strongly %khx
scatter the quasiparticles. The clean regime has been
studied, using a self-consistent solution of the 
Bogoliubov-de Gennes (BdG)\cite{bdg} %khx
equations, in Ref.~\onlinecite{wvh12}. That  work, however,  focused
largely on the transition temperature oscillations. The
results for these oscillations
 were found\cite{exper6} to agree with experiment
 and to be consistent with other %khx
experimentally established\cite{exper1,exper2,exper3,exper5}
results.
In the present work, 
we use the same general 
methods
used there  
to study a simple \sff structure with
noncollinear exchange fields in the ballistic regime, but
we focus on a very different quantity which is readily
accessible experimentally,  namely 
the local DOS and its detailed low-energy structure. 
We strongly emphasize the relation between the ZEP
and the triplet pairing amplitudes. 
In particular, 
given the assertion\cite{exper6}
that 
variations in the transition temperature in these valve structures 
are quantitatively related
to the average triplet pair
amplitudes in the outer $F$ layer, we will search for,
(and, as will be seen, find) correlations between the ZEP and these averages. %khx
This BdG study is complemented with a briefer 
investigation of the corresponding diffusive case.
By considering both regimes, we will  
be able to
provide some general
guidelines for future experiments.

%paragraph split out and edited
The structure we study is
schematically depicted in 
Fig.~\ref{fig:model},
where the 
STM tip is positioned at the outermost $F$ layer, near the vacuum boundary.
In the ballistic regime, we employ the full microscopic  BdG equations
within a self-consistent framework.
From the  solutions,
we calculate the local DOS over a broad range of experimentally 
relevant parameters,
and study its behavior at low energies. 
For the diffusive regime,
we make use of the quasiclassical
Usadel\cite{Usadel} approach to study the 
diffusive \sff counterparts in the
full proximity limit. 
Our systematic investigations thus provide a
comprehensive guide into such spin valves. 
Utilizing experimentally realistic parameters, 
we determine favorable thicknesses for the \f layers to 
induce maximal ZEPs, which 
occurs when the population of triplet correlations in the
outer layer, %otvx2
dominates the singlets.

The paper is organized as follows. In Sec.~\ref{sec:theor} we outline the 
theoretical approaches used. 
In Sec.~\ref{sec:results}, we present our results in two subsections,
pertaining to the ballistic and diffusive regimes. In the ballistic case, 
we study the local 
DOS for differing
exchange field misalignments, exchange field intensities, and interface 
scattering strengths. 
We also investigate the singlet and triplet pairing correlations for 
similar parameters to determine how
ZEPs in the DOS correlate with the triplet correlations. 
In the diffusive case, we present
two-dimensional maps of the ZEP at different exchange field
misalignments, and \fs interface opacity.
Finally, we summarize with concluding remarks in Sec.~\ref{sec:conclusion}.

\section{methods and theoretical techniques} \label{sec:theor}

In this section, we first discuss the theoretical framework used to study
clean samples. We then outline the Usadel technique in the
full proximity regime, which properly describes dirty samples.

\subsection{Microscopic approach: Bogoliubov-de Gennes
equation}\label{subsec:theor_bdg}
For the ballistic regime,
we use the microscopic BdG
equations to study \sff spin valve nanostructures. We solve %otv
these equations in a fully self-consistent\cite{halter1,wvh12} manner.
A schematic of the spin valve
configuration is depicted in Fig.~\ref{model}.
The general spin-dependent BdG equations
for the quasiparticle energies, $\epsilon_n$,
and quasiparticle amplitudes, $u_{n\sigma}$, $v_{n\sigma}$ is written:
\begin{align}
&\begin{pmatrix}
{H}_0 -h_z&-h_x+ih_y&0&\Delta \\
-h_x-ih_y&{H}_0 +h_z&\Delta&0 \\
0&\Delta^*&-({H}_0 -h_z)&-h_x-ih_y \\
\Delta^*&0&-h_x+ih_y&-({H}_0+h_z) \\
\end{pmatrix}
\begin{pmatrix}
u_{n\uparrow}\\u_{n\downarrow}\\v_{n\uparrow}\\v_{n\downarrow}
\end{pmatrix}
\nonumber \\
&=\epsilon_n
\begin{pmatrix}
u_{n\uparrow}\\u_{n\downarrow}\\v_{n\uparrow}\\v_{n\downarrow}
\end{pmatrix},
\label{bogo}
\end{align}
where
the pair potential, $\Delta({x})$, is  calculated 
self-consistently as explained below.
This quasi one-dimensional system
is described by the
single particle Hamiltonian ${\cal H}_0(x)$ as,
\begin{equation}
{\cal H}_0(x)\equiv\frac{1}{2m}\left(-\partial_x^2+k_y^2+k_z^2
\right)-E_F + U(x),
\end{equation}
where $E_F$ is the Fermi energy, and $U(x)$ is the spin-independent interface
scattering potential which we take to be of the form
$U(x)=H [\delta(x-d_{F1})+\delta(x-d_{F1}-d_{F2})]$. %otv H
The in-plane wavevector components, $k_y$ and $k_z$,
arise from the translational invariance in the $y$ and $z$
directions. The system is finite in the $x$ direction, with widths
of each $F$ and $S$ layer shown in the schematic.
Our method permits arbitrary orientations and magnitudes of the magnetic
exchange fields, $\vec {h}_i$ ($i=1,2$), in each of the  ferromagnet regions.
Specifically,
we fix the exchange field in $F_2$ to be aligned
in the $z$ direction,
while in $F_1$, its orientation is described by the angle $\beta_1$:
\begin{align}
\vec{h} = \begin{cases} \vec{h}_{1}=h_0(0,\sin\beta_{1},\cos\beta_{1}), &\mbox{in } F_1 \\
\vec{h}_{2}=h_0 \hat {\bm z}, & \mbox{in } F_2,\end{cases}
\end{align}
where we consider the experimentally appropriate situation of an in-plane
Stoner-type exchange field interaction.

The spin-splitting effects of the exchange field
coupled with the pairing interaction in the $S$ regions,
results in a nontrivial
spatial dependence of the pair potential $\Delta(x)$. In general,
it is necessary  to calculate the pair potential in a self
consistent manner by an appropriate sum over states:
\begin{align} \label{sc}
\Delta(x) = \frac{g(x)}{2}{\sum_{n}}
[u_{n\uparrow}(x)v^*_{n\downarrow} (x)+
u_{n\downarrow}(x)v^*_{n\uparrow} (x)]\tanh\left(\frac{\epsilon_n}{2T}\right),
\end{align}
where $g(x)$ is the attractive interaction that exists solely
inside the superconducting region and  the sum is restricted to
those quantum states with positive energies below
an energy cutoff, $\omega_D$.

We  now discuss the appropriate quantities that  characterize the
induced triplet correlations. We define\cite{halter_trip,halter_trip2}
 the following
triplet pair amplitude functions in terms of the field operators in
the Heisenberg picture,
\begin{subequations}
\label{pa}
\begin{align}
&&{f_0}({x},t) = \frac{1}{2}\left[\left\langle
\psi_{\uparrow}({x},t) \psi_{\downarrow} ({x},0)\right\rangle+
\left\langle \psi_{\downarrow}({x},t) \psi_{\uparrow} ({x},0)\right\rangle\right],~~~~\\
&&{f_1}({x},t) = \frac{1}{2}\left[\left\langle
\psi_{\uparrow}({x},t) \psi_{\uparrow} ({x},0)\right\rangle
-\left\langle \psi_{\downarrow}({x},t) \psi_{\downarrow} ({x},0)\right\rangle\right],~~~~
\end{align}
\end{subequations}
where $t$ is the relative time.
With the quantization axis aligned along the $z$ direction, the time-dependent
triplet amplitudes, $f_{0}(x,t)$ and $f_{1}(x,t)$, can be written in terms of
the quasiparticle amplitudes:~\cite{halter_trip,halter_trip2}
\begin{align}
f_{0}(x,t) &=  \frac{1}{2}\sum_{n}\left(f_n^{\uparrow\downarrow}(x)-f_n^{\downarrow\uparrow}(x)\right) \zeta_n(t), \label{f0} \\
f_{1}(x,t) &  =\frac{1}{2} \sum_{n}
\left(f_n^{\uparrow\uparrow}(x)+f_n^{\downarrow\downarrow}(x)\right)\zeta_n(t), \label{f1}
\end{align}
where we define $f_{n}^{\sigma\sigma'}(x) = u_{n \sigma}(x) v^{\ast}_{n\sigma'}(x)$,
and the time factor $\zeta_n(t)$ is written,
\begin{align}
 \zeta_n(t) = \cos(\epsilon_n t)-i\sin(\epsilon_n t) \tanh\left(\frac{\epsilon_n}{2 T}\right). %khr fixed equation
\end{align}

Experimentally
accessible information regarding the quasiparticle spectra is contained in
the local density of
one particle excitations in the system.
This includes
the zero-energy signatures in the density of states (DOS),
which
present a possible experimental avenue in which to
detect  the emergence of equal-spin triplet correlations within the
outer
ferromagnet.
The total DOS, $N(x,\epsilon)$, is the sum ${N}_\uparrow(x,\epsilon)+{N}_\downarrow(x,\epsilon)$,
involving the spin-resolved local density of states (DOS), $N_\sigma$,
which are written,
\begin{equation}\label{dos}
{N}_\sigma(x,\epsilon)
=-\sum_{n}
\Bigl\lbrace[u^\sigma_n(x)]^2
 f'(\epsilon-\epsilon_n)
+[v^\sigma_n(x)]^2
 f'(\epsilon+\epsilon_n)\Bigr\rbrace,
\end{equation}
where $\sigma$ denotes the spin ($=\uparrow,\downarrow$), and
$f'(\epsilon) = \partial f/\partial \epsilon$ is the derivative
of the Fermi function.

\subsection{Quasiclassical approach: Usadel equation}\label{subsec:theor_usa}

When the system contains a strong impurity concentration, then
for sufficiently small energy scales, the  %otv
superconducting correlations are governed by the Usadel equation.
%where the system contains strong impurities. 
Following
Ref. \onlinecite{DOS_SFS2_th}, the Usadel equation\cite{Usadel}
compactly reads:
\begin{eqnarray}\label{eq:full_Usadel}
&&{\cal D}\Big[\partial,G({\bm r},\epsilon)\big[\partial,G({\bm r},\epsilon)\big]\Big]+\nonumber\\&& i\Big[
\epsilon \rho_{3}+
\text{diag}\big[{\bm h}({\bm r})\cdot\bm\sigma,\big({\bm h}({\bm r})\cdot\bm\sigma\big)^{\cal T}\big],G({\bm r},\epsilon)\Big]=0,
\end{eqnarray}
in which $\rho_{3}$ and $\bm{\sigma}=\big(\sigma^x,\sigma^y,\sigma^z\big)$ are $4\times 4$
and $2\times 2$ Pauli matrices, respectively, and
${\cal D}$ represents the diffusive constant of the magnetic region.
The quasiclassical approach employed in this section supports
ferromagnets with arbitrary exchange field directions;
${\bm h}({\bm r})=\big(h^x({\bm r}),h^y({\bm r}),h^z({\bm r})\big)$.
In Eq. (\ref{eq:full_Usadel}), $G$ represents the total Green's
function which is made of Advanced ($A$), Retarded ($R$), and
Keldysh ($K$) blocks. Therefore, the total Green's function can be
expressed by:
\begin{align}
G({\bm r},\epsilon)=\begin{pmatrix}
G^R & G^K\\
0 & G^A\\
\end{pmatrix},\;\;\;\;G^{R}({\bm r},\epsilon)=\begin{pmatrix}
{\cal G} & {\cal F}\\
-{\cal F}^\ast & -{\cal G}^\ast\\
\end{pmatrix}.
\end{align}
In the presence of ferromagnetism, the components of advanced block,
$G^{A}({\bm r})$, of total Green's function
$G$ can be written as:
\begin{align}
{\cal F}({\bm r},\epsilon)=\begin{pmatrix}
f_{\upuparrows} & f_{\uparrow\downarrow}\\
f_{\downarrow\uparrow} & f_{\downdownarrows} \\
\end{pmatrix},\;\;\;\;{\cal G}({\bm r},\epsilon)=\begin{pmatrix}
g_{\upuparrows} & g_{\uparrow\downarrow}\\
g_{\downarrow\uparrow} & g_{\downdownarrows} \\
\end{pmatrix}.
\end{align}
In this paper,
however, we assume stationary conditions for our systems under
consideration, and hence %kh
the three blocks comprising the total Green's function are
related to each other in the following way:
$G^{A}({\bm r},\epsilon)=-\big[\rho_3 G^R({\bm r},\epsilon)\rho_3\big]^{\dag}$, and
$G^{K}({\bm r},\epsilon)=\tanh(\beta\epsilon)\big[G^{R}({\bm r},\epsilon)-G^{A}({\bm r},\epsilon)\big]$, where $\beta\equiv k_BT/2$.

The \fs interface controls the proximity effect. Therefore,
appropriate boundary conditions should be considered to properly
model the system. In our work, we consider the Kupriyanov-Lukichev boundary conditions at
the \fs interface\cite{cite:zaitsev} which controls the induced
proximity correlations using a parameter $\zeta$ as the barrier
resistance:
\begin{eqnarray}\label{eq:bc}
    &&\zeta G({\bm r},\epsilon)\partial G({\bm r},\epsilon)=
    [G_{\text{BCS}}(\theta,\epsilon),G({\bm r},\epsilon)].
\end{eqnarray}
The solution for a bulk
even-frequency $s$-wave superconductor $G_{\text{BCS}}^{R}$
reads,\cite{MK_jap}
\begin{align}
\hat{G}^{R}_{\text{BCS}}(\theta,\epsilon)=\left(
                                  \begin{array}{cc}
                                    \mathbf{1}\cosh\vartheta(\epsilon) & i\sigma^y \sinh\vartheta(\epsilon) \\
                                    i\sigma^y \sinh\vartheta(\epsilon) & -\mathbf{1}\cosh\vartheta(\epsilon) \\
                                  \end{array}
                                \right),
\end{align}
where
$\vartheta(\epsilon)=\text{arctanh}(\mid\Delta\mid/\epsilon)$.

The system local density of states, ${\cal N}({\bm r},\epsilon)$,
can be expressed by the following equation:
\begin{equation}
{\cal N}({\bm r},\epsilon)=\frac{{\cal N}_0}{2}\text{Re}
\Big[\text{Tr}\big\{G({\bm r},\epsilon)\big\}\Big],
\end{equation}
in which ${\cal N}_0$ is the density of state normal state.

\section{Results and discussion} \label{sec:results}

In this section, we describe our results. We start with those for a ballistic
\sff structure and then present the predictions of Usadel formalism for 
diffusive samples.

\subsection{Ballistic Regime}\label{subsec:res_cln}
In this subsection we 
present the self-consistent results
for the ballistic regime.
The numerical method  used here to iteratively solve
in a self consistent way
Eqs.~(\ref{bogo}) and (\ref{sc}) has been  extensively described  
elsewhere,~\cite{halter1,wvh12} and details need not be repeated here.
In the calculations, the temperature $T$ is held constant at $T=0.05 T_c$,
where $T_c$ is the transition temperature of a pure bulk $S$
sample.
All length scales are normalized  by the 
Fermi wavevector, so that
the coordinate $x$ is written $X = k_F x$,
and the $F_1$ and $F_2$ widths are written
$D_{Fi}\equiv k_F d_{Fi}$, for $i=1,2$.
The  ferromagnet $F_2$ and superconductor  are set to fixed values,
corresponding to
 $D_{F2} = 400$,
and  $D_S=600$, respectively.
We  also assume a coherence length corresponding to $k_F \xi_0 = 100$.
One of our main objectives in this paper is to study the triplet correlations,
which are odd in time.\cite{halter_trip}
To accomplish  this, we employ the expressions  in  Eqs.~(\ref{f0}) and (\ref{f1}), 
which describe the  
spatial and temporal behavior of 
the triplet amplitudes.
At $t=0$  the triplet correlations 
vanish because of the Pauli
exclusion principle. 
At finite $t$, the triplet correlations
generated  near the $S/F$ interface tend to increase in
amplitude and spread throughout the structure.
We normalize the time  $t$
% in these expressions is normalized
according to $\tau=\omega_D t$, and
we set it to a representative\cite{wvh12} value
of $\tau=4$.
We can  then 
study the   
behavior of
the triplet amplitudes $f_0$ and $f_1$ throughout  the junction.
To explore the proximity induced signatures in the single-particle states,
which is the main purpose of this work,
we then present a systematic investigation of
the experimentally relevant local DOS.
All DOS results presented are 
local values taken at a fixed position near %otv precisely?
the edge of the sample in the $F_2$ region.
We characterize  interface scattering, when  present,
by delta functions of strength $H$,
which we write in terms of the
 dimensionless parameter $H_B\equiv H/v_F$.
Finally, we use natural units, e.g., $\hbar=k_B=1$ throughout.

\subsubsection{Triplet and singlet pair correlations}\label{trip}
Here we present results for  both the triplet and singlet correlations,
calculated using Eqs.~(\ref{f0})-(\ref{f1}). 
For the cases shown  below,  
the absolute value of the singlet and triplet complex quantities 
are  averaged over the region of interest, which in this case
is the  experimentally probed $F_2$ region. An important reason
for focusing on those spatially averaged  (over the
outer magnet) quantities, rather than the spatial
profiles discussed in Ref.~\onlinecite{wvh12},
 is that it was experimentally shown\cite{exper6} that these 
triplet averages perfectly anticorrelate with the transition 
temperatures, i.e. the spin valve  effect.    %otv
We also normalize all pair correlations to the value of the bulk
singlet pair amplitude. 
\begin{figure}[]
\includegraphics[width=0.4\paperwidth]{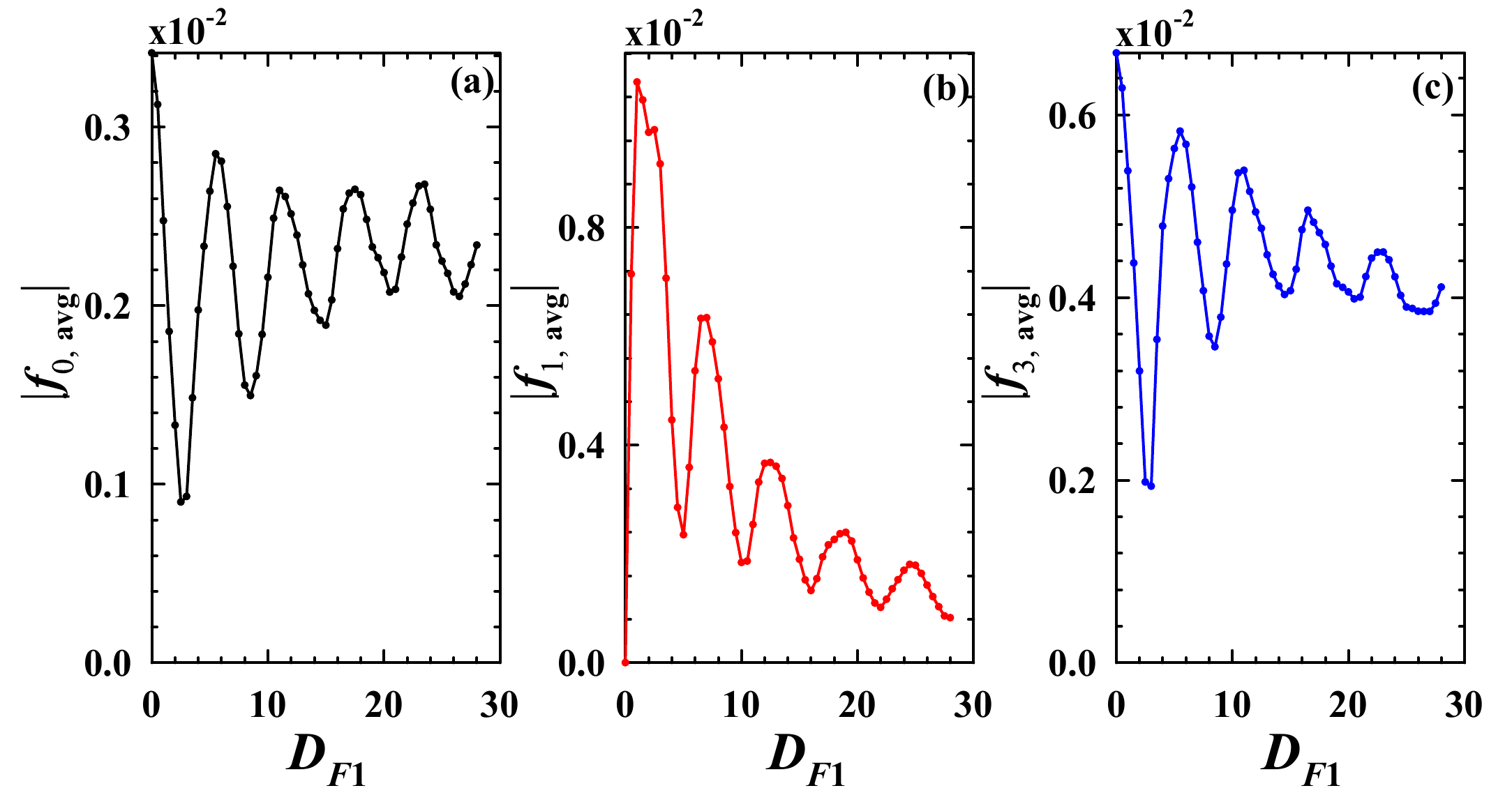}
\includegraphics[width=0.4\paperwidth]{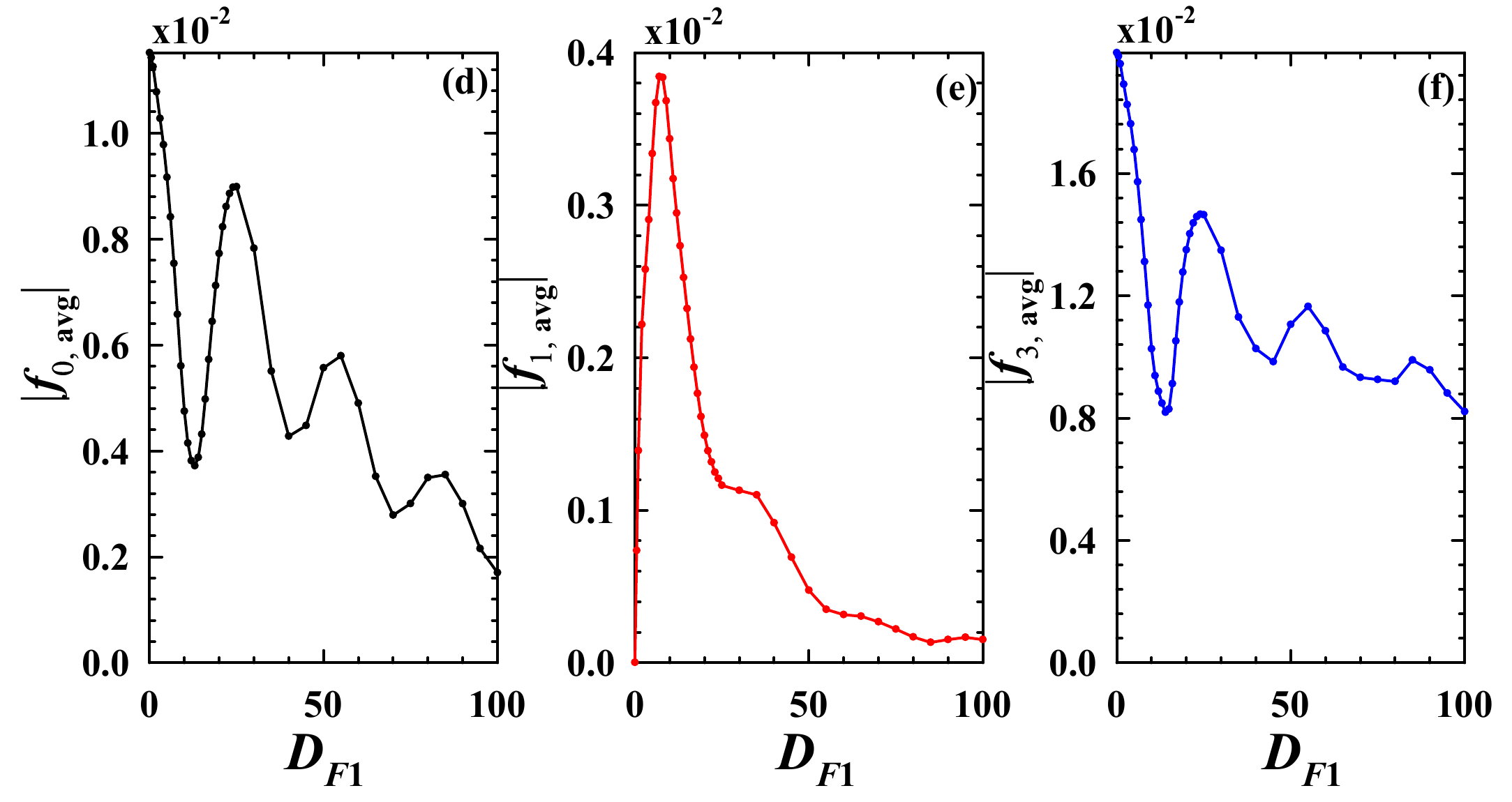}
\includegraphics[width=0.4\paperwidth]{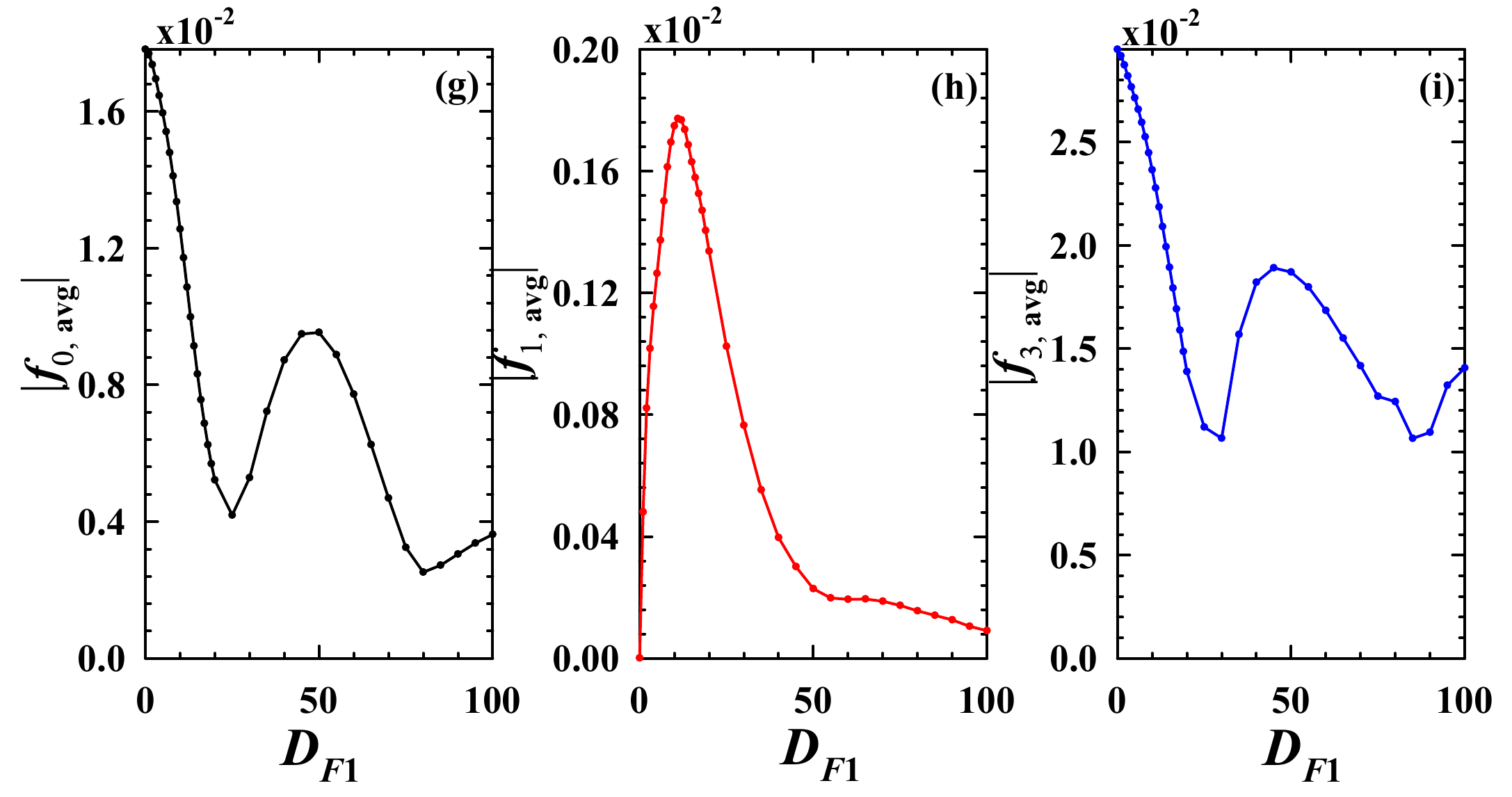}
\caption{(Color online) 
The absolute value of the normalized triplet and singlet pair
correlations, averaged over the $F_2$ region, 
as a function of $D_{F1}$. 
% For each quantity the magnitude is taken and  averaged over the $F_2$ region.
The exchange field strengths are (from top to bottom):
$h/\varepsilon_F=0.5,0.1,0.05$.
The relative exchange field orientations
are orthogonal, with
$\beta_1=\pi/2$,
and $\beta_2=0$.
}
\label{trip_df1}
\end{figure}
We begin by showing, in Fig.~\ref{trip_df1},
the spatially averaged absolute value
of the complex triplet amplitudes $|f_{0,{\rm avg}}|$ (with spin projection $m=0$),
and $|f_{1,{\rm avg}}|$ (with spin projection $m=\pm1$) along with the
singlet $|f_{3,{\rm avg}}|$, (note that %otv undefined
$f_3(x) \equiv \Delta(x)/g(x)$) as functions of $D_{F1}$.
Each row of panels corresponds to a different exchange field value:
from top to bottom  rows, we have $h/\varepsilon_F=0.5,0.1$, and $0.05$.
Examining the opposite spin correlations, $f_0$ and $f_3$, damped oscillatory
behavior with $D_{F1}$ is evident: this is related to the spatial oscillation %khx
of the Cooper pairs due to their acquiring a center of mass momentum when entering the
magnet.~\cite{demler}
Therefore, the wavelength of these oscillations varies inversely with the exchange field in $F_2$
(this is why the $D_{F1}$ range for the weaker exchange fields is extended).
Quantum interference effects
generate
peaks in $f_0$ and $f_3$ that 
occur  approximately  when $d_{F1}/\xi_F=n\pi$,
%or equivalently, 
(i.e. $D_{F1}=n\pi (\he)^{-1}$).
In the ballistic regime, the
length scale  that characterizes the damped oscillations is  $\xi_F=v_F/ (2h)$,
where $v_F$ is the Fermi velocity.
The equal-spin amplitudes $f_1$, are seen to behave oppositely, with a phase offset of
 approximately $\pi/2$. 
Their magnitude % of the $f_1$ amplitudes also 
declines more rapidly with
$D_{F1}$, compared to the behaviors of $f_0$ and $f_3$.
This is consistent with $f_1$ triplet generation being optimal for
highly asymmetric ferromagnetic layer widths.~\cite{MK_jap}
It is notable that 
the periodic occurrence of peaks in $f_1$ when varying 
$D_{F1}$, evolves into a single maximum as $h$ is reduced further.

One of the strengths of the microscopic BdG formalism is 
having the
ability to properly include the full microscopic range of length
and energy scales inherent to
the problem. This includes the exchange energy $h$, which
in our BdG framework 
can span the limits from a nonmagnetic normal metal ($h/\varepsilon_F=0$)
to a half-metallic ferromagnet ($h/\varepsilon_F=1$).
It is particularly useful to consider  the behavior of the singlet and triplet correlations 
over this  broad range of  strengths of $h/\varepsilon_F$.
\begin{figure}[]
\includegraphics[width=0.4\paperwidth]{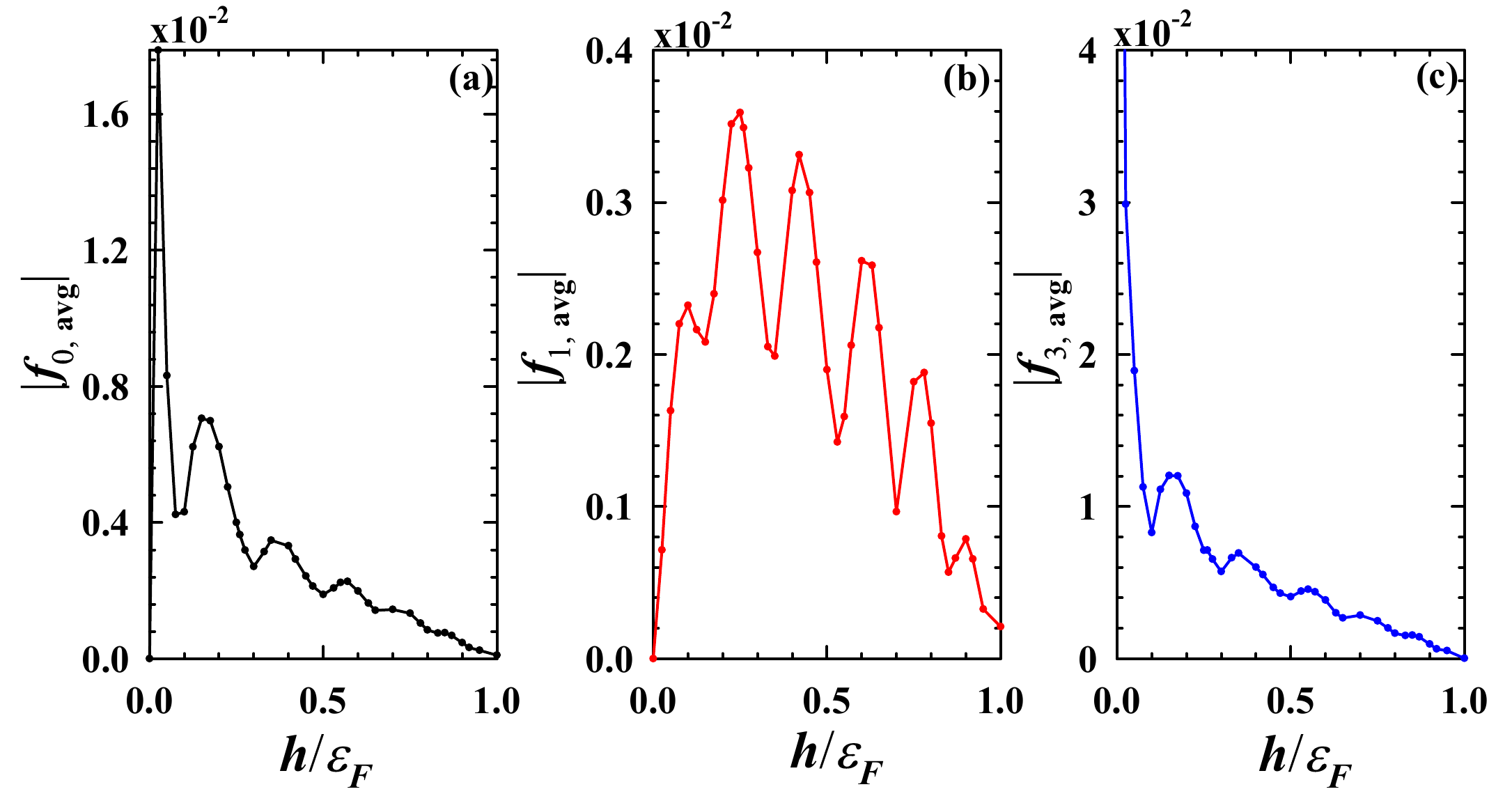}
\includegraphics[width=0.4\paperwidth]{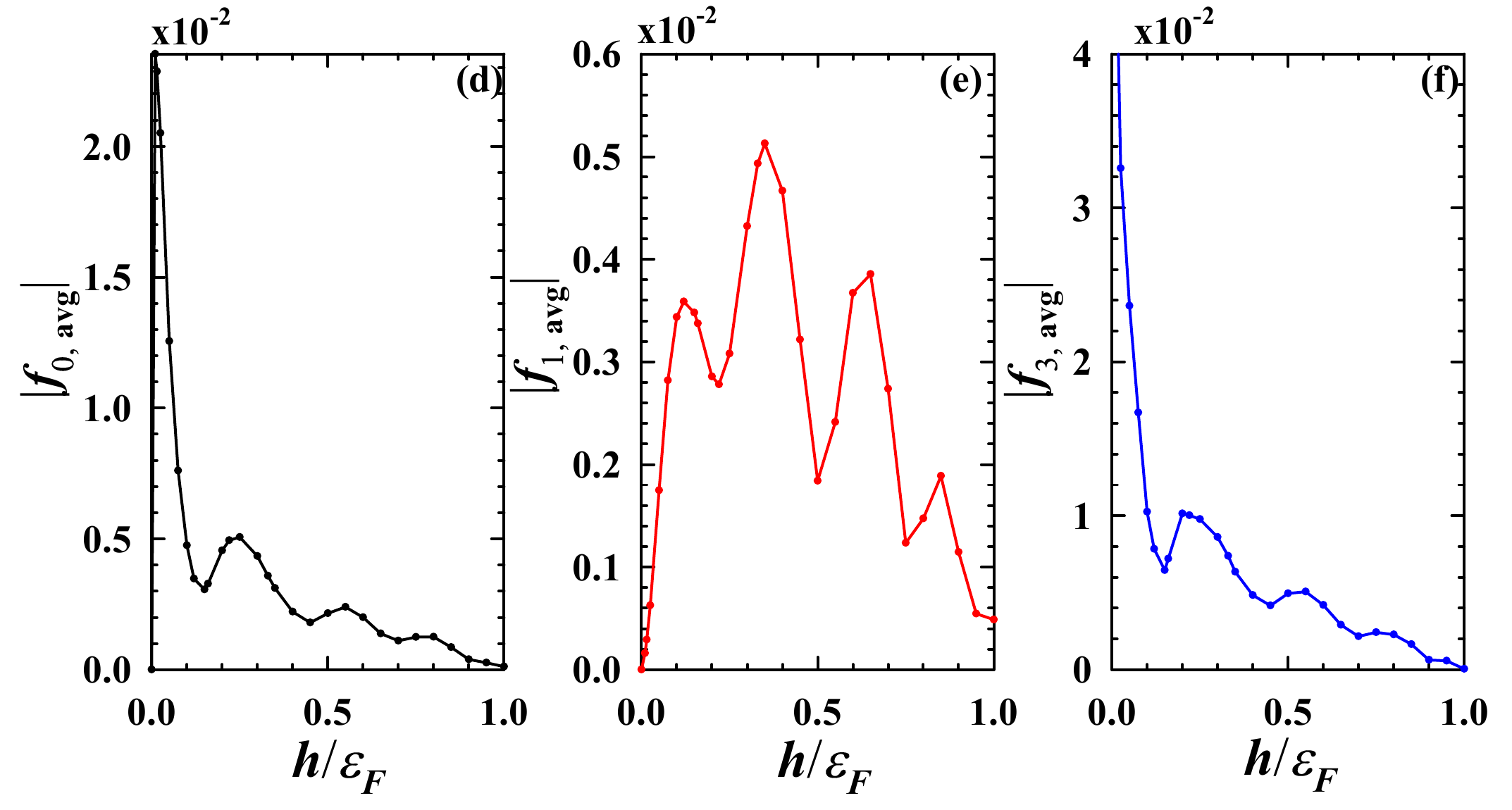}
\includegraphics[width=0.4\paperwidth]{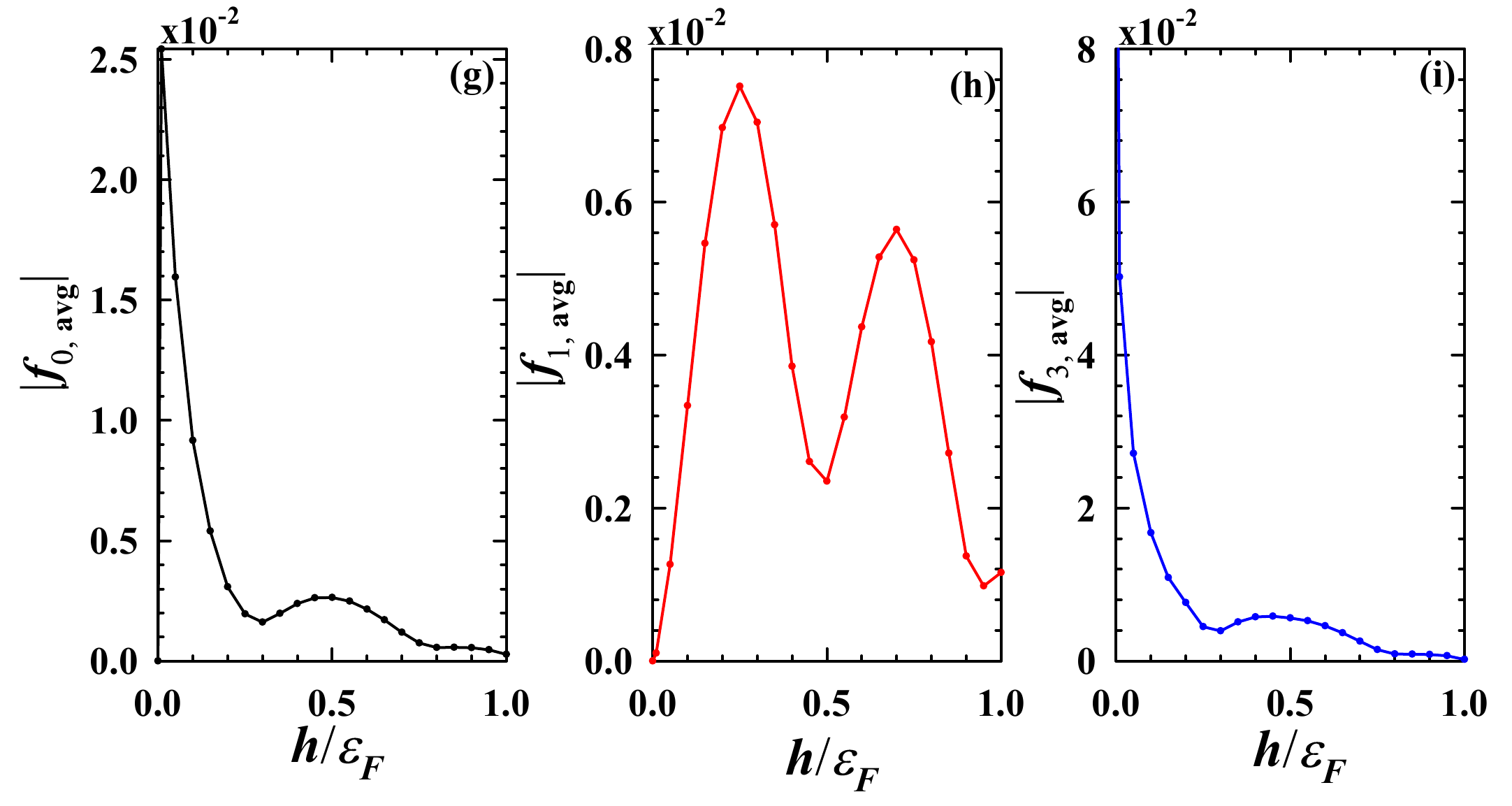}
\caption{(Color online) 
The spatially averaged (in $F_2$) normalized  triplet and singlet pair
correlations 
as a function of $h/\varepsilon_F$. 
As in Fig.~\ref{trip_df1},  the magnitude 
of each quantity is taken and  averaged over the $F_2$ region.
Each row of panels corresponds to a different
$F_1$ width, with
$D_{F1}=15$ (top row), $D_{F1}=10$ (middle row),
and $D_{F1}=5$ (bottom row).
The relative exchange field orientations
are orthogonal, with
$\beta_1=\pi/2$,
and $\beta_2=0$.
}
\label{trip_h}
\end{figure}
Thus, in  Fig.~\ref{trip_h}, we 
show the same quantities as Fig.~\ref{trip_df1},
plotted now as a function  $h/\varepsilon_F$.
Again, we have
orthogonal  relative exchange field orientations, with
$\beta_1=\pi/2$,
and $\beta_2=0$.
Each three-panel row  corresponds to a different $F_1$ width:
 $D_{F1}=15,10$, and $5$ (from top to bottom).
 The central column reveals
 that the averaged equal spin amplitudes $|f_{1,{\rm avg}}|$ displays regularly occurring prominent peaks,
 the number of which varies with the length of the $F_1$ region.
For the exchange fields and $F_1$ widths considered in Fig.~\ref{trip_df1},
the triplet $f_1$ was generally weaker than either the singlet $f_3$ or triplet $f_0$. 
For the system parameters used in Fig.~\ref{trip_h} however,
we find that
for narrow widths $D_{F1}$
and sufficiently
large exchange fields, 
the equal-spin triplet component  $f_1$ can  dominate the
other pair correlations. 
In particular, for strong 
ferromagnets with $h/\varepsilon_F\approx 0.8$,
and  thin $F_1$ layers with $D_{F1}=5$,
panels (g) and   (i) illustrate that the $f_0$ and $f_3$ amplitudes
 consisting of
opposite spin pairs,  
are negligible due to the pair breaking effects of the strong
magnet.
On the other hand, the equal-spin pairs shown in panel (h)
are seen to survive in this limit.
This 
has important consequences for isolating 
and measuring this triplet
component in experiments.

Having seen how the magnitude of the exchange field $h$
affects the  singlet and triplet correlations,
we next 
 investigate
 the effects of changing its direction.
\begin{figure}[t!]
\includegraphics[width=0.4\paperwidth]{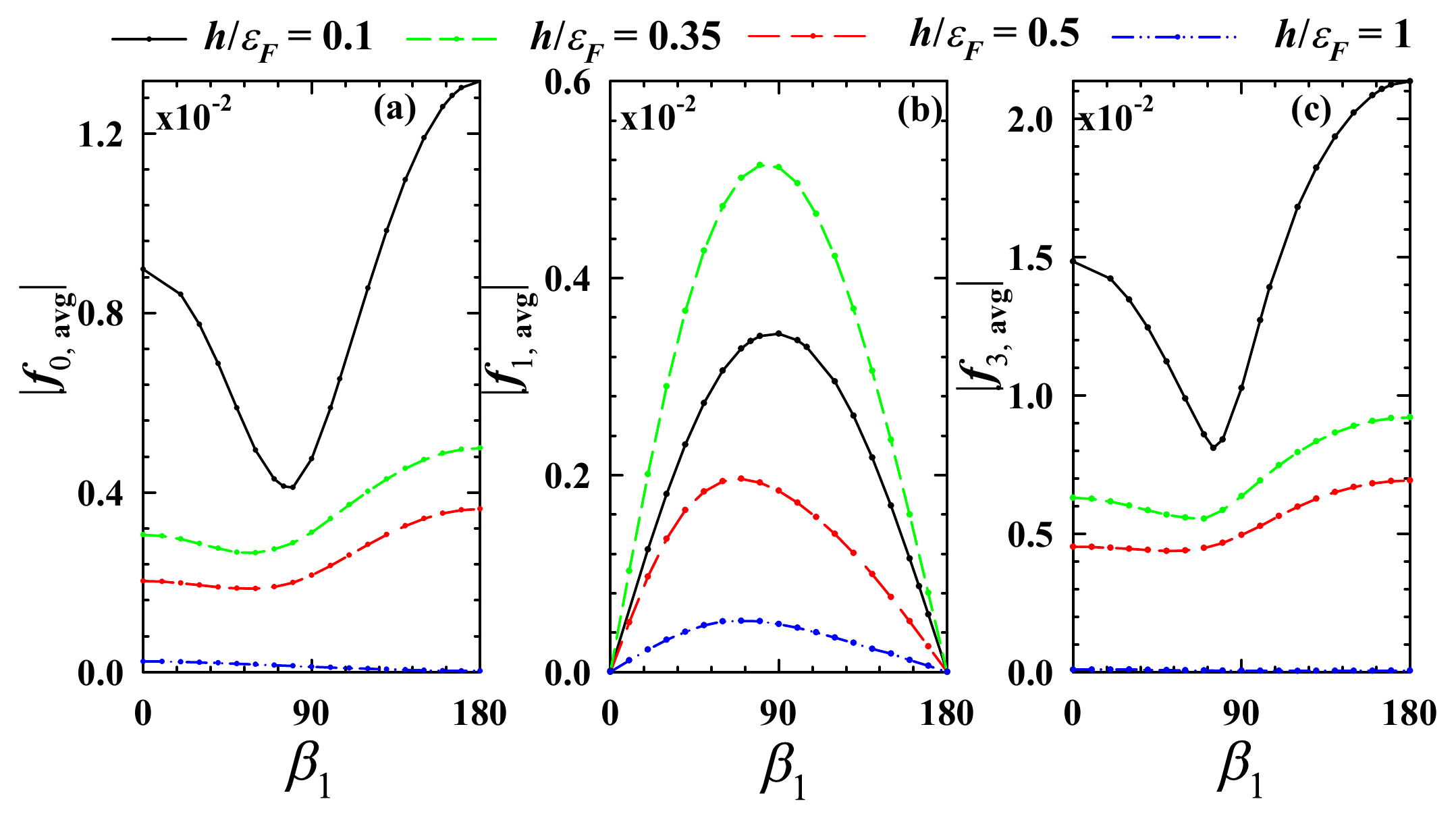}
\caption{(Color online) Plots of the
averaged singlet and triplet components as a function of magnetic orientation $\beta_1$.
Here $D_{F1} = 10$,
and results for several magnetic strengths are shown, ranging from weak to half-metallic.
}
\label{trip_phi1}
\end{figure}
Therefore, 
we examine in 
Fig.~\ref{trip_phi1}, the behavior of the
averaged singlet and triplet amplitudes
when changing the magnetic orientation angle, $\beta_1$.
We again consider a broad range of exchange field strengths, as shown in the legend.
One of the more obvious features is that
the maximum of $|f_{1,{\rm avg}}|$ typically does not  
occur for orthogonal relative exchange fields,~\cite{karmin}  %khx
for smaller $\beta_1\lesssim 90^\circ$,
especially for stronger magnets. 
This is in agreement with previous\cite{fominov,karmin,ah,exper6,wvh12,yu,colci}
experimental and theoretical results. %otv
Due to the non-monotonicity  of  $|f_{1,{\rm avg}}|$ with $h$ [see Fig.~\ref{trip_h}(e)],
the $h/\varepsilon_F=0.35$ case seen in Fig.~\ref{trip_phi1}(b) %khx
is larger 
for all $\beta_1$  than for the  weaker $h/\varepsilon_F=0.1$ case.
The singlet $f_3$ and triplet 
$f_0$ amplitudes are largest 
for antiparallel 
configurations 
($\beta_1=180^\circ$), where the opposite exchange fields %khx degrees, not radians, to be consistent?
are
effectively weakened,
with reduced spin-splitting effects on the  
opposite-spin Cooper pairs. This is a well-known result.
The results also show that the
relative magnetic orientation 
angles leading to the minima of these two quantities are
anticorrelated with
the angles at which the  $f_1$ correlations are maximal. 
As seen in Figs.~\ref{trip_phi1}(a) and (c), 
$|f_{0,{\rm avg}}|$ and $|f_{3,{\rm avg}}|$  decay much more abruptly 
as the value of $h$ in the magnets approaches the half-metallic limit: 
this is consistent with the discussions above. Therefore, \sff structures
involving  strong ferromagnets ($h\sim \varepsilon_F$)
with $\beta_1$   at or near orthogonal orientations,
can host larger generated triplet
pair correlations whereby $|f_1|\gg
\{|f_0|, |f_3|\}$, thus allowing for direct probing of  the spin triplet
superconducting correlations in experiments.

\begin{figure}[]
\includegraphics[width=0.4\paperwidth]{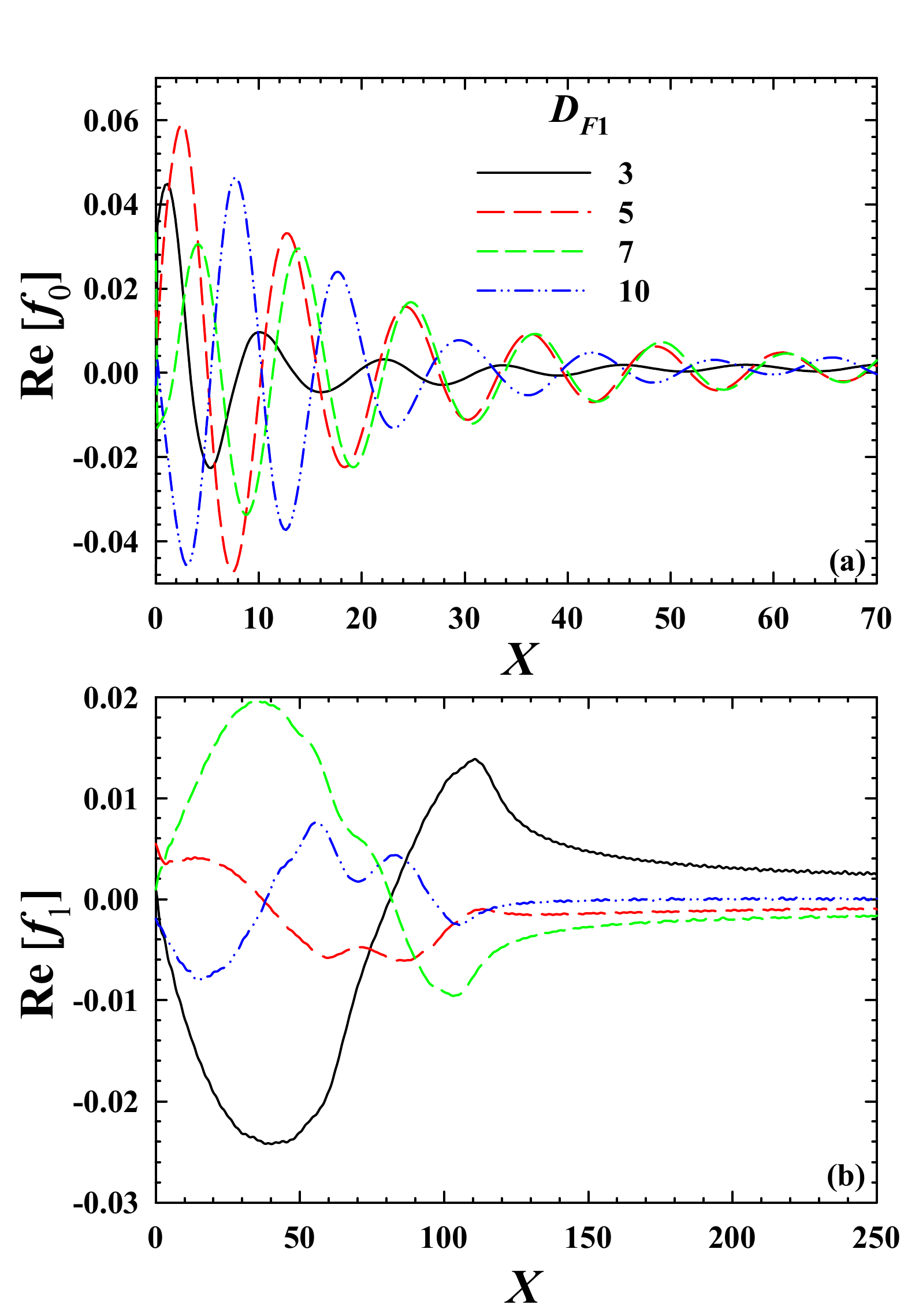}
\caption{(Color online) Local spatial profiles of the real parts of the triplet components
$f_0$ and $f_1$ in  the $F_2$
region for a few different $F_1$ widths, $D_{F1}$.
The exchange field in the ferromagnets corresponds to $\he =0.5$,
and the  relative exchange field orientations are orthogonal, with
$\beta_1=\pi/2$,
and $\beta_2=0$.
}
\label{trip_x}
\end{figure}
More detailed information regarding the triplet amplitudes,
can be obtained from  the spatial profiles of the local triplet correlations 
within the $F_2$ region.
In Fig.~\ref{trip_x}, we present  the real
parts of the normalized $f_0(x)$ and $f_1(x)$ triplet components 
in terms of the dimensionless coordinate $X$.
Results are plotted at four different values
of $D_{F1}$ as indicated in the legend.
The exchange fields in the ferromagnets has magnitude corresponding
to $\he=0.5$, and the directions are mutually orthogonal, with $\beta_1= 90^\circ$, and $\beta_2=0$.
For
the time scale considered here, the imaginary part of $f_0$
is typically much smaller than its real part. As to $f_1$, its imaginary part is
usually not negligible, but it exhibits trends that are similar to those for the real part.
Examining the top panel, it is evident that %otv no labels
 $f_0$ exhibits the trademark damped oscillatory
spatial dependence
arising from the difference in the spin-up and spin-down wavevectors
of the Cooper pairs.
The  oscillatory wavelength is thus governed by the quantity 
	$2\pi k_F \xi_F = 2\pi (h/\varepsilon_F)^{-1}$, which
	for our parameters corresponds to  $4\pi$.
	The modulating $f_0$ has the same
	wavelength for each  $D_{F1}$,  
	although each curve can differ in phase.
	The averaged $f_0$ amplitudes  
	are consistent with this local behavior:  Fig.~\ref{trip_df1}(a)
	demonstrated that 
	when $D_{F1}\approx 5$ and $D_{F1}\approx10$,  there is an enhancement of the $f_0$ component,
	while for $D_{F1}\approx 3$, it is substantially reduced.
The equal-spin $f_1$ amplitudes,  are shown in the bottom panel of %otv
Fig.~\ref{trip_x}. Near the interface at $X=0$, the $f_1$ correlations  are created,
 and then they subsequently increase in magnitude  until deeper within the ferromagnet,
where  they
clearly exhibit a gradual long-ranged decay.
The  trends observed here are opposite to those in the top panel, where for instance
the $D_{F1}=3$ case leads to maximal $f_1$ triplet generation, in agreement with 
Fig.~\ref{trip_df1}(b).

\subsubsection{Local density of states} \label{dos_sec}
After the discussion of the salient
features  of the singlet and triplet  pair correlations in the outer $F$ layer,  %khx
%otv that reside throughout the $SFF$ structure, 
we now turn to the main topic of the
paper: the local
density of states measured in $F_2$.  
This is the experimentally relevant quantity %otv 
that can reveal the signatures of these correlations.
The damped oscillatory behavior of 
the pair correlations
can lead to spectroscopic 
signatures in the form of DOS 
inversions,~\cite{kontos}  and multiple oscillations.~\cite{DOS_norm_SF1}
In the  quasiclassical approximation,~\cite{fominov,karmin,yu,ah} %otv $SFF$ systems,
a
ZEP can emerge 
from the
long-range triplet correlations~\cite{Kawabata,MK_jap} in $SFF$ systems.
However, 
this approximation is not %otvx2 be 
appropriate for
experimental conditions
involving strong  magnets and clean interfaces. %otv
It would be beneficial experimentally  to 
characterize  the ZEP  
 relation to
the singlet and triplet correlations and see how
the ZEP may be a useful fingerprint  in 
identifying 
the existence of the long-range triplet component. %otv
To properly do this over the broad range of 
parameters found in experimental conditions,
a microscopic self-consistent theory
 that
can accommodate the  wide ranging length and energy scales 
is needed.
In this subsection, we therefore present  an extensive 
microscopic study
of  the ZEP
%of the 
%and how that width depends on various
as a function of
parameters such as $F$ layer thicknesses, exchange energy, or
interface transparency. 
These results are then
correlated with the self-consistent 
singlet and triplet pair correlations in the previous subsection.
In what follows, the DOS is normalized by
the DOS at the Fermi level ${\cal N}_F$, and
plotted vs the normalized energy $\varepsilon/\Delta_0$,
where $\Delta_0$ is the bulk value of the pure $S$
material gap at zero temperature.
Our emphasis will be on energies within the subgap 
region $\varepsilon\leq \Delta_0$,
where the  ZEP phenomenon arises.
Since the DOS is a local quantity that depends on position [see Eq.~(\ref{dos})],
in our calculations we assume the location to be near the edge of the sample
just below the STM tip as shown in Fig.~\ref{fig:model}.

\begin{figure}[t!]
\includegraphics[width=0.4\paperwidth]{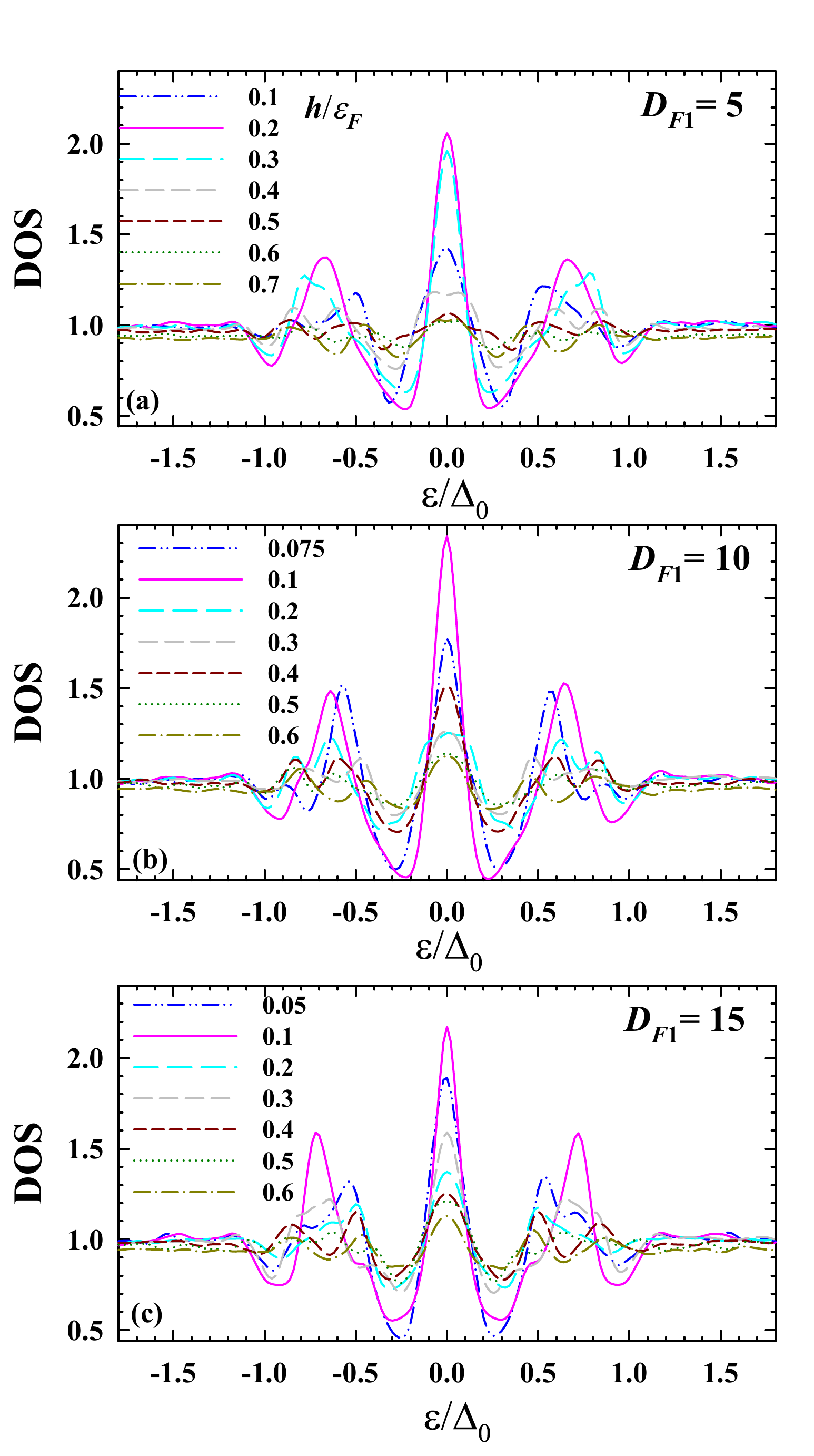}
\caption{(Color online)
Normalized (see text) local DOS. The\
 multiple curves in each panel are for
different values of $h/\varepsilon_F$. 
Each panel corresponds to a different value of $D_{F1}$ (as labeled).
The ferromagnets have exchange fields with
orthogonal relative directions. %khx
}
\label{dos_df1}
\end{figure}
To correlate the 
triplet 
correlations in 
Fig.~\ref{trip_h}
with the ZEP,
we begin by studying in Fig.~\ref{dos_df1} 
the sensitivity of the DOS to a broad range of 
exchange field strengths $h$.
Each panel corresponds to a different $F_1$ 
width, $D_{F1}$.
The 
range of $h$ 
considered in 
each panel varies since the largest ZEP 
depends on the relative values of $h$ 
and $D_{F1}$. 
The top panel ($D_{F1}=5$) 
clearly shows the progression of the ZEP with $h$:
Beginning with the smallest exchange field, $\he = 0.1$,
 a moderate peak is observed that increases to its maximum height
and narrower  width when $\he=0.2$. Further increases in $h$
 continuously diminish the ZEP,  broadening its width, until eventually 
 it is effectively washed away.
This non-monotonic behavior % where the ZEP
% is suppressed at $\he=0.1$ compared to $\he=0.3$,
is consistent with the ZEP being related to 
the presence of the $f_1$ triplet amplitude near the edge of  the ferromagnet.
This can be seen by reexamining  the
triplet amplitudes in Fig.~\ref{trip_h}(h), 
where the exchange field  leading to the highest ZEP occurs when
$|f_{1,{\rm avg}}|$ is largest, at $\he \approx 0.2$.
%otv BUT secondary peak at 0.7 is missing!
The same consistency is found between Figs.~\ref{trip_df1}(e) and (b)
and the middle and lower panels of Fig.~\ref{trip_h}, respectively. 
For both the $D_{F1}=10,15$ cases, the average value of
$|f_1|$ is largest 
near $\he = 0.1$. However, the secondary peak structure  in 
Fig.~\ref{trip_df1} is not clearly refected in the DOS. %otv?

\begin{figure}[t!]
\includegraphics[width=0.4\paperwidth]{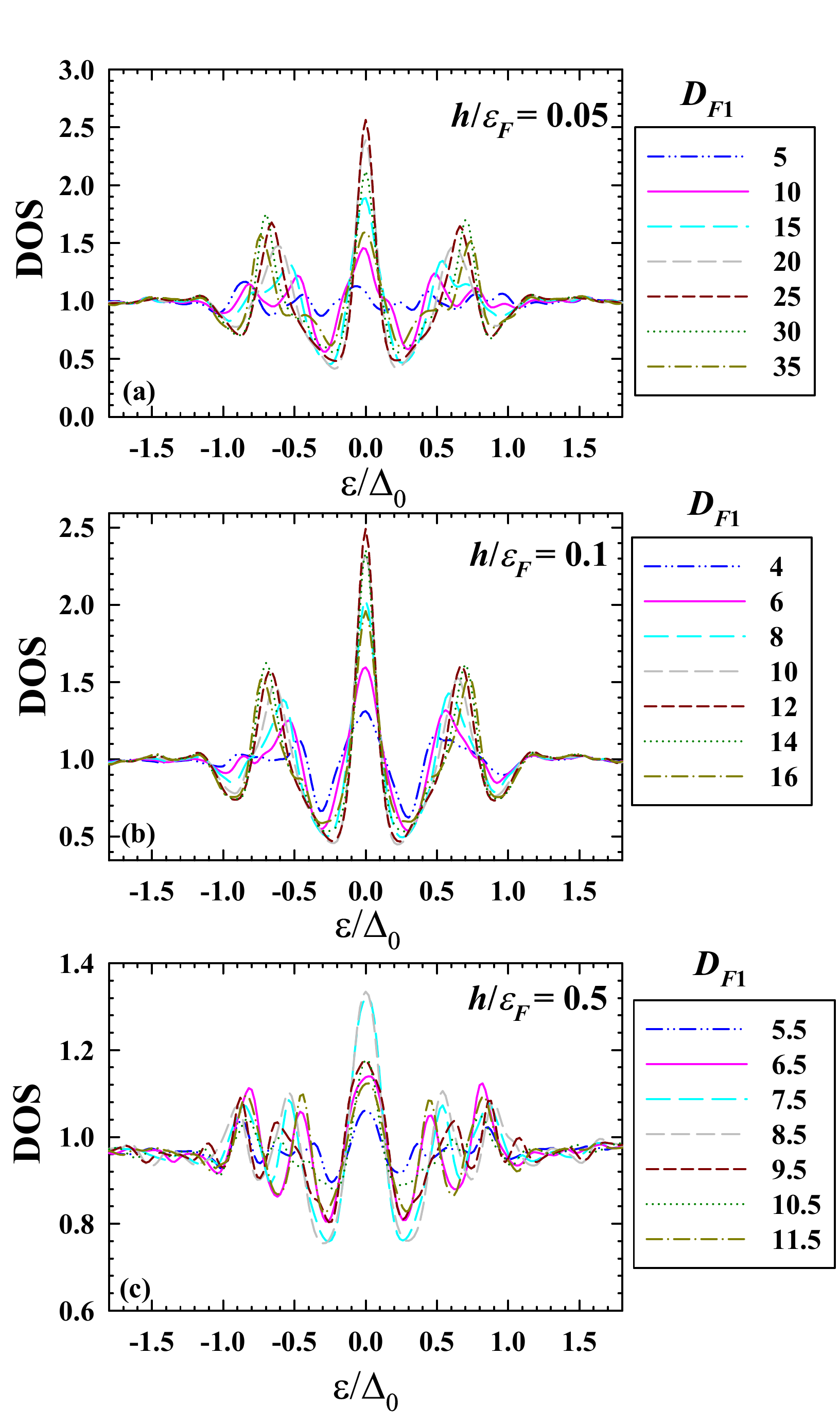}
\caption{(Color online)
Normalized local DOS as a function of the normalized energy. 
The  curves in each panel are for 
different values of the width $D_{F1}$. 
Each panel corresponds to a different  
$\he$: $\he=0.05, 0.1, 0.5$,
and
we consider  orthogonal relative exchange fields.
}
\label{dos_df2}
\end{figure}
Next we study the DOS
counterpart to Fig.~\ref{trip_df1}.
The
normalized DOS, and the corresponding
ZEP, are shown in Fig.~\ref{dos_df2}
for a broad range of widths $D_{F1}$.
The
parameter values 
here are similar to 
those used in Fig.~\ref{trip_df1}, where each
panel corresponds to 
a different exchange field.
In panel (a) with $\he=0.05$, the most prominent ZEP occurs for
$D_{F1}=25$, coinciding with the $F_1$ width that yields a local maximum 
for the $m=0$ triplet amplitude $f_0$
(see Fig.~\ref{trip_df1}(g)).  
By comparison, the $f_1$ component  observed in Fig.~\ref{trip_df1}(h),
 is smaller and lacks the multiple peak structure found
for $f_0$, at this weaker exchange field.
Therefore, the largest ZEP in the case of weak exchange fields,
does not necessarily occur when the triplet $f_1$ is maximal;
as Fig.~\ref{trip_df1}(h) demonstrated,  $|f_{1,{\rm avg}}|$ peaks at $D_{F1}=10$
before  rapidly declining. For these weaker fields, it follows from
Fig.~\ref{trip_df1} that the magnitude of $f_0$ exceeds that of $f_1$. %otv
It would appear then that it is the larger triplet component which %khx
determines the ZEP structure. This is consistent with the %otv
known result\cite{exper6} that  the {\it total} value of the triplet component
is correlated  with $T_c$.
The next case in panel (b) 
corresponds also to 
a moderately weak magnet with
$\he=0.1$,
or double the exchange field considered in panel (a).
Since the frequency of the oscillations involving 
the opposite-spin $f_0$ amplitudes [see Fig.~\ref{trip_df1}(d)]
also doubles, the maximum ZEP at $D_{F1}=12$,
occurs at about half the $F_1$ width found 
for the maximum ZEP in (a).
The equal-spin triplet correlations $f_1$ 
were seen in  Fig.~\ref{trip_df1}(e)
to exhibit a single peak structure, 
but their magnitude is larger than at weaker
fields. This is because
typically
stronger magnets in this situation lead to an enhancement of the 
$f_1$ amplitudes. 
 Lastly, we consider (bottom panel)
 a relatively strong ferromagnet with $\he=0.5$.
 For this case, there are additional subgap peaks flanking the main ZEP.
 The larger ZEP arises at smaller widths ($D_{F1}=7.5,8.5$) than 
 for weaker exchange fields,
 due to an increase in the frequency of the oscillations as a function
of $D_{F1}$ for the $f_0$ and $f_1$ components as
 seen in Fig.~\ref{trip_df1}(a) and (b).
 Thus, the  ZEP tends to exhibit a structure 
 that dampens and widens  for strong magnets,
 while the opposite is true for weaker ones and is correlated with the
 stronger of the $m=0$ and $m=\pm 1$ triplet components present. %otv

\begin{figure}[t!]
\includegraphics[width=0.4\paperwidth]{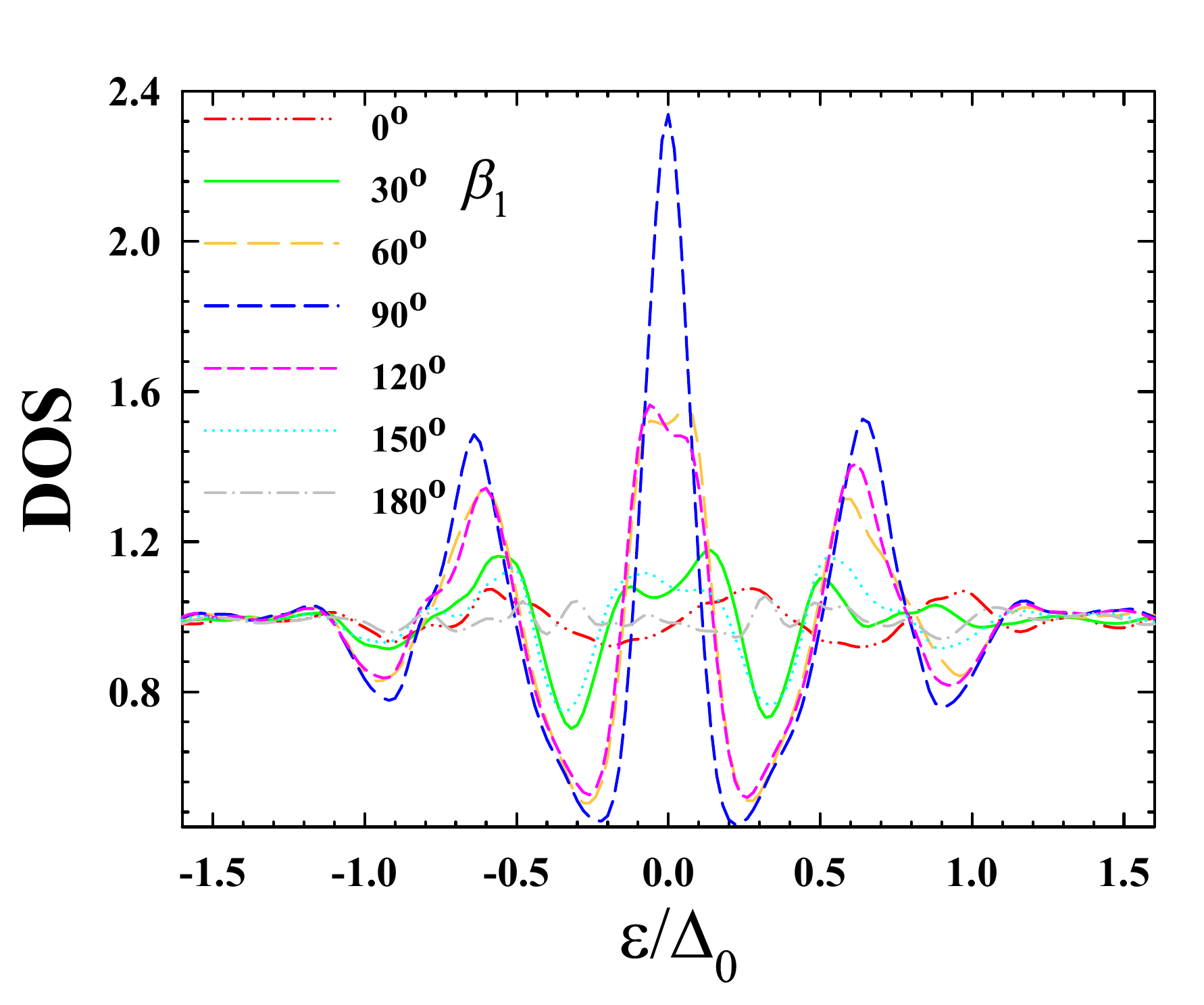}
\caption{(Color online)
Variation of the normalized local DOS with the in-plane exchange field angle 
$\beta_1$. The exchange field is fixed along $z$ in $F_2$.
Also, 
$D_{F1}=10$, and
 $\he=0.1$.
}
\label{dos_phi1}
\end{figure}
Having established the behavior of the ZEP for differing $\he$, 
we now fix the magnitude
of the exchange fields in each magnet and
investigate the effects of varying their relative orientation. 
Figure~\ref{dos_phi1} illustrates the normalized DOS
for the specific case $D_{F1}=10$, and
%specifically
 $\he=0.1$. According to
Fig.~\ref{trip_phi1}(b), the equal-spin triplet 
component $f_1$
is greatest  when  $\beta_1\approx 90^\circ$. 
Thus we would expect the ZEP to also be maximal at 
this angle. Figure~\ref{dos_phi1} shows that this is indeed the case. 
There the normalized DOS
is shown for a range of $0^\circ\leq \beta_1\leq 180^\circ$ in increments of
 $30^\circ$. 
 Clearly the orthogonal 
 relative exchange field ($\beta_1=90^\circ$)
 configuration results in the
 most prominent ZEP.  When $\beta_1$ deviates from this angle
 towards the P ($\beta_1=0^\circ$) or AP  ($\beta_1=180^\circ$) 
 alignments,
 both the triplet amplitude $f_1$ and the ZEP decline until
  $\beta_1=0^\circ$ or $180^\circ$, whereby $f_1=0$, 
  and the ZEP has vanished.

\begin{figure}[]
\includegraphics[width=0.4\paperwidth]{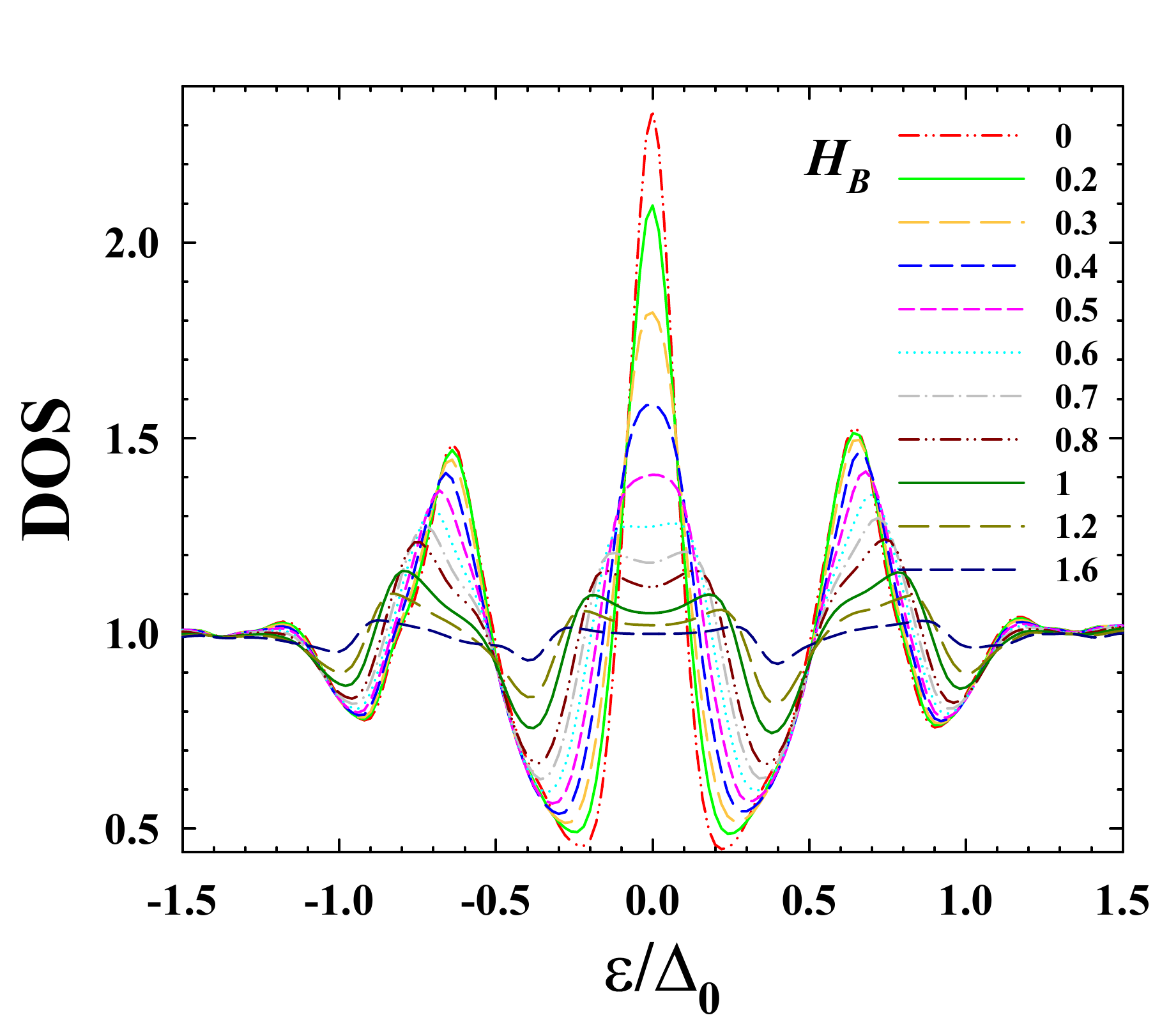}
\caption{(Color online)
Evolution of the ZEP with scattering strength $H_B$:
The normalized local DOS is shown as a function of the dimensionless energy. 
Each curve depicts results for a
different  scattering strength $H_B$ (see text).
The system parameters correspond to 
$D_{F1}=10$, and $\he=0.1$.
The exchange fields in the ferromagnets are mutually orthogonal with
$\beta_1=90^\circ$,
and $\beta_2=0^\circ$.
\label{dos_hb}
}
\end{figure}

Finally, in Fig.~\ref{dos_hb}
we examine the effects 
of interface scattering on
the self-consistent energy spectra.
We assume that each interface has the same delta function
potential barrier
with dimensionless scattering strength $H_B$.
We consider a broad range of $H_B$, from transparent interfaces with $H_B=0$,
to very high interfacial  scattering, with $H_B=1.6$.
By allowing $H_B$ to vary, we effectively control the  proximity effects:
a small $H_B$ results in  stronger proximity coupling between the 
$F$ and $S$ regions,
while a large $H_B$ results in isolation of each segment, 
and weak proximity effects.
This is evident in the DOS, as seen in  Fig.~\ref{dos_hb}, 
which has its largest ZEP when $H_B=0$.
The width and height of the ZEP is 
strongly influenced by the presence of interface scattering.
Increasing $H_B$ results in the ZEP widening while gradually 
diminishing in height.
Eventually, when the scattering strength reaches 
$H_B\approx 0.7$, the peak begins to split.
Further increments in $H_B$
causes the  peaks to  separate 
and eventually proximity effects are so weakened 
that the DOS becomes that of  an isolated bulk ferromagnet.
The two secondary subgap peaks that lie symmetrically about the ZEP
are seen to also decline in a monotonic fashion as $H_B$ becomes larger.

\subsection{Diffusive Regime}\label{subsec:res_diff}
In this section, we consider a diffusive \sff junction in the full
proximity limit. We employ the Usadel approach described in
Sec.~\ref{sec:theor} to  investigate the local DOS. 
As remarked earlier, the quasiclassical method is limited
to energies close to the Fermi level. Hence, our discussion
here will be  limited to
relatively weak ferromagnets.
As in the
ballistic regime, we consider %otv moved 
heterostructures where the magnetic layers 
are made of identical materials so that 
the ferromagnetic coherence lengths are the same, 
$\xi_{F_1}=\xi_{F_2}\equiv \xi_F$,
and we consider the low temperature regime where $T =0.05 T_c$. 
Prior to calculating the DOS,
we normalize the Usadel equation  by 
$\xi_F$,
which in the diffusive regime is written, %otv ???
$\xi_F=\sqrt{D/h}$. Using this normalization scheme, the explicit
dependency on the exchange field is removed and
the Usadel equation now involves terms containing the ratio
$d_F/\xi_F$. %khr which F? %otv  the same see below.
This approach  can lead 
to easier pinpointing of regions in parameter space where
the ZEP is most prominent, and
it also permits
a broad range of this ratio to be studied. 
We assume
that the magnetic orientation angle is fixed  
at $\beta_2=0$,
or equivalently ${\bm h}=(0,0,h_z)$. %and %$T =0.05T_c$. 

\begin{SCfigure*}
  \centering
\includegraphics[width=12.0cm,height=6.20cm]{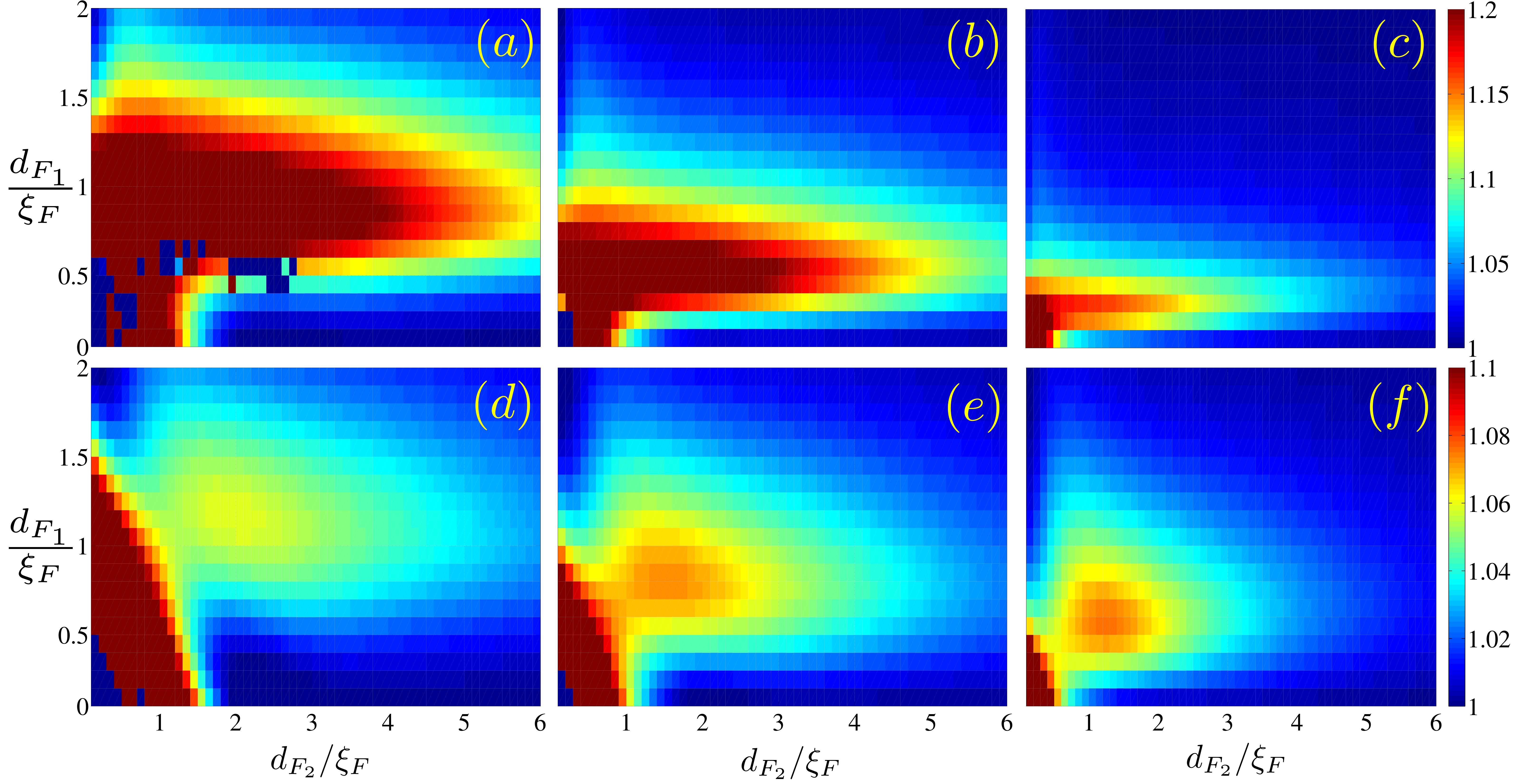}
\caption{\label{diff_2d} (Color online) 
Zero energy peak in the DOS spectrum of diffusive \sff spin valves as a 
function of the normalized \f layer
thicknesses $d_{F1}/\xi_F$
and $d_{F2}/\xi_F$.
Each column corresponds to a different \fs transparency, 
with $\zeta=1, 2.5, 5$ (from left to right).
The top row of panels shows the evolution of the 
ZEP for the misalignment angle $\beta_1=\pi/2$, while the bottom
row of panels are for 
$\beta_1=\pi/6$. 
For both cases, the internal field of the $F_2$ layer is  
along the $z$ direction, $\beta_2=0$.
The ZEPs are computed at $x=d_{F1}+d_{F2}$ 
(at the top most \f /vacuum interface).
}
\end{SCfigure*}
We numerically solve the Usadel equation, Eq.~(\ref{eq:full_Usadel}),
together with the mentioned boundary conditions. 
To find the total Green's function, %otvx2 did not look right
we substitute the solution into Eq.~(\ref{eq:full_Usadel}) and obtain the DOS.
To determine the optimal geometry in which the ZEPs are most pronounced,
we present in
Fig.~\ref{diff_2d} 
the ZEP at the topmost edge of the \sff structure,
corresponding to the location
$x=d_{F1}+d_{F2}$.
The two-dimensional color mapping depicts the 
strength of 
the DOS at zero energy (the ZEP) 
as a function of the
normalized $F$ thicknesses,  $d_{F1}/\xi_F$ and  $d_{F2}/\xi_F$. 
In the top row 
panels, the internal field 
of the \f layers have a misalignment angle
 of $\beta_1= \pi/2$, while for the bottom row
$\beta_1=\pi/6$. 
The left, middle, and right columns  
are for
different opacities at the \fs
interface: $\zeta=1.0$,
$2.5$, and $5.0$, respectively. 
By increasing  $\zeta$,
the overall strength of the  proximity effects
is   effectively 
weakened: 
it is evident that transparent \fs contacts yield 
stronger ZEPs, that persist in thicker \f layers. 
It is also 
apparent that the orthogonal case $\beta_1=\pi/2$ 
has more extensive regions in the  parameter space
spanned by the $F$ thicknesses
with enhanced
ZEPs, as  compared to  the $\beta_1=\pi/6$ case.
An important aspect of the ZEP  
that all cases investigated in Fig.~\ref{diff_2d}
share, is that it is strongest when $d_{F1} \ll d_{F2}$. 
This finding is fully consistent with low proximity 
 bilayer SFF hybrids.\cite{ah}
Therefore,  for
the parameters considered here, the
ZEPs are strongest for $\zeta=1$, $0.5\xi_F \lesssim d_{F1} \lesssim \xi_F$,
and $1.5\xi_F \lesssim d_{F2} \lesssim 3.5\xi_F$. 
The ratio of the $F$ thickness to the length scale $\xi_F$ is an
important dimensionless quantity that appears in the normalized 
Usadel equations,  
and consequently   thinner  $d_{F1}$ and $d_{F2}$,
allow for stronger 
ferromagnets   when studying  the DOS  in the diffusive limit.

\begin{figure}[]
\includegraphics[width=6.50cm,height=6.60cm]{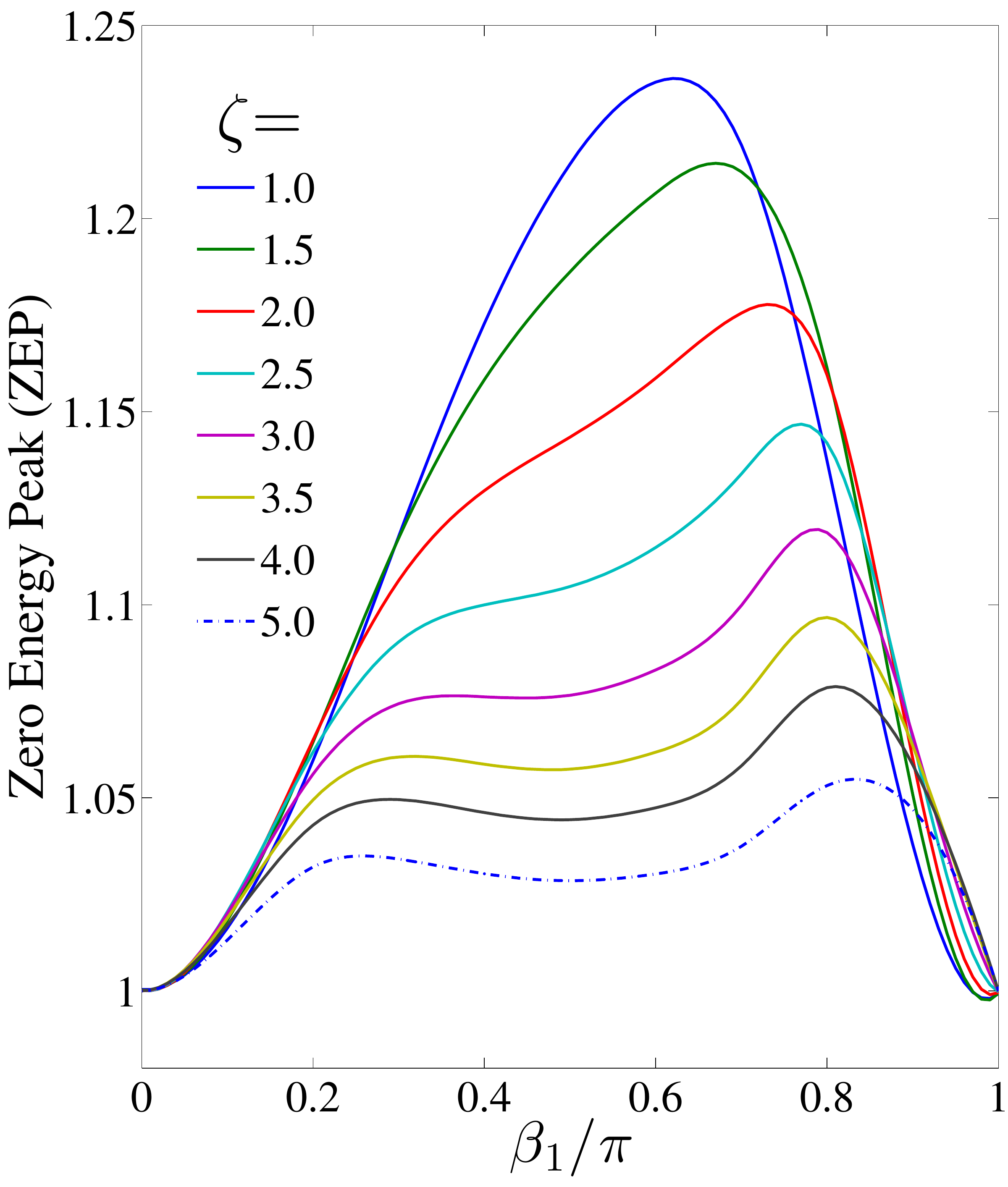}
\caption{\label{diff_1d} (Color online) Zero energy peak of the 
DOS spectrum in diffusive 
\sff spin valves as a function of exchange field orientation $\beta_1$ 
for several values of 
$\zeta$, which controls the opacity of the  \fs interface. 
We set $\beta_2=0$ and rotate the exchange field direction of $F_1$ from the
parallel  ($\beta_1=0$) to antiparallel ($\beta_1=\pi$) orientations. 
We have chosen representative values of $d_{F1}=0.8 \xi_F$ and $d_{F1}=3.5 \xi_F$, 
in accordance with  the system parameters used  in Fig.~\ref{diff_2d}.
 }
\end{figure}
Finally, we study
the sensitivity of the ZEPs to both the orientation angle $\beta_1$ and
interface transparency  parameter $\zeta$.
We thus show in 
Fig.~\ref{diff_1d}
the ZEP as a function of $\beta_1$
over a wide range of  $\zeta$,  as shown in the legend. 
The geometric parameters correspond to  %khx2
$d_{F1}=0.85 \xi_F$ and $d_{F2}=3.5 \xi_F$,
which resides within the range of system widths studied in Fig.~\ref{diff_2d}
resulting in the largest ZEPs.
In calculating the ZEP, 
we  again consider
the DOS at the edge of the sample (see also Fig.~\ref{model}).
It is seen that the interface transparency 
can  significantly 
alter the
behavior of the ZEP as  the
relative exchange field angle  sweeps from the 
P ($\beta_2=0^\circ$) to AP ($\beta_2=180^\circ$)
orientations.
For example, when $\zeta=1.0$, the maximal ZEP 
is offset from the orthogonal configuration,
occurring  at
$\beta_2\approx 0.6\pi$. 
By increasing the barrier strength, this peak
shifts  towards larger $\beta_2$,
until the relative exchange fields are nearly antiparallel.
There is also a simultaneous reduction
in amplitude, due to the $F$ and $S$ regions
becoming decoupled as the  proximity effects diminish. 
Interestingly, as $\zeta$ increases, there is a splitting of the main peak:
weaker secondary peaks emerge. 
Eventually however, for sufficiently large $\zeta$,
the opacity of the interface causes the low energy DOS to be insensitive to 
%variations
$\beta_1$, and the ZEP flattens out. 
The ZEPs are also observed to disappear when
the relative exchange fields are collinear,
corresponding to the situation when the triplet amplitudes vanish
in both the diffusive and ballistic regimes (see also Fig.~\ref{trip_phi1}).

%otvr edited conclusions including title
\section{Summary and Conclusions} \label{sec:conclusion}

In summary, we have employed a microscopic self-consistent wavefunction 
approach to study the low energy
proximity induced local DOS in \sff spin valves with
noncollinear exchange fields in the clean limit. Our emphasis has been
on the results of STM methods that probe the outer $F$ layer. To 
identify the physical source of the corresponding ZEPs that occurs
for such data in these systems,
we also calculated the absolute value  of the triplet    pair
correlations, averaged over the outer $F$ layer. We have
done so for a broad range of experimentally relevant 
parameters, including the exchange field strength and orientation, 
as well as thicknesses of the ferromagnets. Our results demonstrate
a direct link between the spin-1 triplet correlations and  the appearance of 
ZEPs in the local DOS spectra, and 
point to % optimal is too strongoptimal 
system parameters and configurations which 
would support larger %otv careful maximal 
equal-spin triplet superconducting correlations.
These correlations could then be probed indirectly via 
single-particle signatures
that are measurable using local  spectroscopy techniques.
Our results are consistent with\cite{exper6}  findings relating
the average strength of triplet correlations to the angular dependence
of the transition temperature.
Our findings suggest that the ZEPs arising from the spin-1 triplet amplitudes
can be  effectively isolated in  \sff  systems with
strong ferromagntets, with the outer one being
very thin. % magnet, and a  larger one.
This  asymmetric geometry not only produces greater equal-spin triplet
generation, but it  can also filter out the rapidly 
decaying opposite-spin pairs 
deep within the sample.
We also 
considered the same valve structure in the diffusive regime utilizing
a Green function method within the full proximity limit.
Our investigations yielded 
a broad range of
\f layer thicknesses and relative exchange fields orientations
that lead to observable signatures in the
low energy DOS, thus also
 giving useful guidelines for
future experiments.

\acknowledgments

K.H. is supported in part by ONR
and by a grant of supercomputer resources provided by the DOD HPCMP.
We thank N. Birge for many  useful discussions about %otv
the ZEP problem and for providing
thoughtful feedback on a draft of this work.
M.A. would like to thank A.A. Zyuzin for helpful discussions.

\end{document}